\documentclass{arxiv}

\begin{document}

\title{Near-Field Aeroacoustic Shape Optimization at Low Reynolds Numbers}
\date{}
\author{Mohsen Hamedi and Brian C. Vermeire\\
\textit{Department of Mechanical, Industrial, and Aerospace Engineering}\\
\textit{Concordia University} \\
\textit{Montreal, QC, Canada}} 
\maketitle

\begin{abstract}
We investigate the feasibility of gradient-free aeroacoustic shape optimization using the Flux Reconstruction (FR) approach to study two-dimensional flow at low Reynolds numbers. The Overall Sound Pressure Level (OASPL) is computed via the direct acoustic approach, and optimization is performed using the gradient-free Mesh Adaptive Direct Search (MADS) algorithm. The proposed framework is assessed across three problems. First, flow over an open cavity is investigated at a Reynolds number of $Re=1500$ and free-stream Mach number of $M_\infty = 0.15$, resulting in a $7.9dB$ noise reduction. The second case considers tandem cylinders at $Re=200$ and $M_\infty = 0.2$, achieving a $16.5 dB$ noise reduction by optimizing the distance between the cylinders and their diameter ratio. Finally, a NACA0012 airfoil is optimized at $Re=10,000$ and $M_\infty = 0.2$ to reduce trailing edge noise. The airfoil's shape is optimized to generate a new 4-digit NACA airfoil at an appropriate angle of attack to reduce OASPL while maintaining the baseline time-averaged lift coefficient and preventing an increase in the baseline time-averaged drag coefficient. The optimized airfoil is silent at $0dB$ and the drag coefficient is decreased by $24.95\%$. These results demonstrate the feasibility of shape optimization using MADS and FR for aeroacoustic design.
\end{abstract}

\section{Introduction}

The World Health Organization (WHO) estimates a loss of more than a million healthy life years in Western Europe annually due to noise \cite{world2011burden}. 
The negative impacts of noise on both the environment and human health have been well established \cite{mahashabde2011assessing, basner2017aviation}.  Environmental impacts include disruptions to wildlife behavior and habitat \cite{pepper2003review}, while human health impacts can range from hearing loss to cardiovascular disease \cite{basner2017aviation}. 
Air transportation is a major contributor to noise pollution including airframe noise, turbine and compressor noise, combustion noise, and jet noise, amongst others. It is anticipated that the total number of air passengers will double by 2040 \cite{IATAprediction}. It follows that these noise impacts from aviation must be taken into consideration and reduced accordingly.

Accurate aviation noise reduction requires both a precise noise prediction technique and a suitable optimization strategy.  Aeroacoustic challenges, distinct from conventional aerodynamics, arise due to their inherent transient nature.  Firstly, aerodynamic problems are generally dominated by relatively low-frequency momentum-dominated structures \cite{tam2006recent}. However, aeroacoustic problems are dominated by high-frequency and high-velocity acoustic waves traveling over large distances.  These waves typically have minute amplitudes, considerably smaller than the mean flow \cite{tam2006recent}. Thus,  employing a numerical method with minimal dispersion and dissipation errors becomes imperative for accurately computing sound waves. An additional challenge in CAA is the slow decay rate of acoustic waves. This often leads to reflections from computational domain boundaries, thereby contaminating the numerical solution. Recent advancements in high-order numerical techniques, like the Flux Reconstruction (FR) approach \cite{huynh2007flux}, offer promise for CAA applications. These methods exhibit reduced numerical errors compared to conventional low-order Finite Volume Methods (FVM) \cite{moura2015linear, vincent2011insights, vermeire2017behaviour} and are well-suited for modern hardware architectures \cite{vincent2016towards}.
 
Growing aviation demand necessitates the development of next-generation aircraft to be quieter and environmentally sustainable, requiring robust optimization frameworks. There are two main categories of optimization techniques: gradient-based and gradient-free.  Gradient-based methods are efficient for smooth and well-behaved objective functions but struggle with unsteady problems due to challenges in accurately computing gradients in chaotic flows \cite{johnson1995computatbility, chen2019discretization, martins2013multidisciplinary, karbasian2022gradient}. In contrast, gradient-free methods, like Genetic Algorithm (GA) \cite{forrest1996genetic, schmitt2004theory}, Particle Swarm Optimization (PSO) \cite{praveen2009low, wang2011robust, zhang2015comprehensive}, and Mesh Adaptive Direct Search (MADS) \cite{audet2006mesh}, do not require sensitivities, making them suitable for complex and noisy objective functions. These methods depend solely on multiple objective function evaluations to approach the optimal design \cite{karbasian2022gradient}, and are typically more robust with a greater probability of finding the global optimum solution \cite{luo2014aerodynamic, lyu2015aerodynamic}; however, they can become prohibitively expensive when a large number of design parameters are employed \cite{karbasian2022gradient}.  Karbasian \cite{karbasian2022gradient} demonstrated the feasibility of shape optimization using MADS and FR for aerodynamic design, performing an aerodynamic shape optimization on a low Reynolds SD7003 airfoil resulting in a $32\%$ increase in the mean lift-to-drag ratio. Furthermore, Aubry et al. \cite{aubry2022high} performed shape optimization of low-pressure turbine cascade using the MADS optimization algorithm, resulting in significant improvements in performance with reasonable computational cost \cite{aubry2022high}.

In this study, the overall sound pressure level at an observer location is computed. Then, the shape optimization procedure is conducted using the gradient-free MADS technique to reduce noise at this observer. Gradient-free optimization techniques have previously been used to reduce the trailing-edge noise \cite{marsden2007trailing, kholodov2019optimization, rottmayer2023trailing} and the propeller noise \cite{xue2023multidisciplinary, peixun2020aeroacoustic, marinus2010aeroacoustic}. However, to the author's knowledge, there is no previous work on coupling the high-order FR approach to the MADS optimization technique for aeroacoustic problems. Thus, the objective of this study is to investigate the feasibility of high-order numerical techniques coupled with gradient-free optimizers for aeroacoustic problems.

This paper is outlined as follows. The methodology is given in Section \ref{sec:Methodology}. Then, the shape of a two-dimensional open cavity is optimized to reduce noise in Section \ref{sec:DeepCavity}, followed by two-dimensional tandem cylinders in Section \ref{sec:TandemCylinders}, and, airfoil shape optimization for noise reduction is performed in Section \ref{sec:NACA0012}. Finally, the conclusions and recommendations for future work are given in Section \ref{sec:Conclusions}.

\section{Methodology}
\label{sec:Methodology}

This section presents an overview of the methodology employed to solve the unsteady Navier-Stokes equations.

\subsection{Governing Equations}

The compressible unsteady Navier-Stokes equations can be cast in the following general form
\begin{equation}
\frac{\partial \pmb{u}}{\partial t} + \pmb{\nabla} \cdot \pmb{F} = 0,
\label{equation_conservation_law}
\end{equation}
where $t$ is time and $\pmb{u}$ is a vector of conserved variables
\begin{equation}
\pmb{u} = 
\begin{bmatrix}
\rho \\
\rho u_i \\
\rho E
\end{bmatrix},
\end{equation}
where $\rho$ is density, $\rho u_i$ is a component of the momentum, $u_i$ are velocity components, and $\rho E$ is the total energy. The inviscid and viscous Navier-Stokes fluxes are 
\begin{equation}
\pmb{F}_{i,j} (\pmb{u}) = 
\begin{bmatrix}
\rho u_j \\
\rho u_i u_j + \delta_{ij} p \\
u_j ( \rho E + p)
\end{bmatrix},
\end{equation}
and
\begin{equation}
\pmb{F}_{\nu, j} (\pmb{u}, \nabla \pmb{u}) =
\begin{bmatrix}
0 \\
\tau_{ij} \\
-q_j - u_i \tau_{ij}
\end{bmatrix} ,
\end{equation}
respectively, where $\delta_{ij}$ is the Kronecker delta. The pressure is determined via the ideal gas law as
\begin{equation}
p = (\gamma - 1) \rho \left( E - \frac{1}{2} u_k u_k \right) ,
\end{equation}
where $\gamma=1.4$ is the ratio of the specific heat at constant pressure, $c_p$, to the specific heat at constant volume, $c_v$. The viscous stress tensor is
\begin{equation}
\tau_{ij} = \mu \left( \frac{\partial u_i}{\partial x_j} + \frac{\partial u_j}{\partial x_i} - \frac{2}{3} \frac{\partial u_k}{\partial x_k} \delta_{ij} \right) ,
\end{equation}
and, the heat flux is
\begin{equation}
q_j = - \frac{\mu}{Pr} \frac{\partial}{\partial x_j} \left( E + \frac{p}{\rho} - \frac{1}{2} u_k u_k \right) ,
\end{equation}
where $\mu$ is the dynamic viscosity and $Pr=0.71$ is the Prandtl number. 

\subsection{Flux Reconstruction}

The FR approach \cite{huynh2007flux}, discretizes the divergence operator for general advection-diffusion equations of the form shown in Eq. (\ref{equation_conservation_law}). Known for its high-order accuracy, generality, robustness, and compatibility with modern hardware architectures \cite{vincent2016towards}, FR offers superior accuracy with fewer degrees of freedom and reduced computational cost compared to conventional low-order methods \cite{vermeire2017utility}. Its suitability extends to scale-resolving simulations, where its numerical error behavior is leveraged for ILES \cite{vermeire2016implicit}, and via filtering approaches for highly under-resolved problems \cite{hamedi2022optimized}. The FR framework is outlined here in multiple dimensions, following Wang's formulation \cite{wang2009unifying}.

In the FR approach, the computational domain, $\Omega$, is discretized into a mesh of $N_e$ non-overlapping elements such that
\begin{equation}
\Omega = \bigcup_{k=1}^{N_e} \Omega_k , \quad \quad \bigcap_{k=1}^{N_e} \Omega_k = \emptyset.
\end{equation}
Each element, $\Omega_k$, contains a number of solution points based on the desired solution polynomial degree. For the sake of simplicity, these elements are transformed from the physical space $\pmb{x}$ to a standard reference space $\pmb{\xi}$, where $\pmb{x}$ and $\pmb{\xi}$ are the spatial coordinates in the physical and reference spaces, respectively. The transformation of these elements is done via an invertible mapping function, $M$, such that
\begin{equation}
\pmb{x} = M \left( \pmb{\xi} \right) \quad \Longleftrightarrow \quad \pmb{\xi} = M^{-1} \left( \pmb{x} \right).
\end{equation}
The Jacobian of this mapping can be found at any point from
\begin{equation}
J = \frac{\partial \pmb{x}}{\partial \pmb{\xi}},
\end{equation}
which enables all element operations to be performed on the same reference element and, upon completion, mapped back to the physical locations.
 
\begin{figure}
\centering
\begin{subfigure}{0.35\textwidth}
\includegraphics[width=\textwidth]{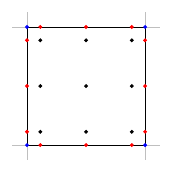}
\subcaption{Quadrilateral element types.}
\end{subfigure}
\begin{subfigure}{0.35\textwidth}
\includegraphics[width=\textwidth]{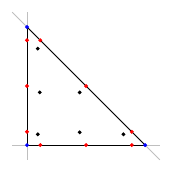}
\subcaption{Triangular element types.}
\end{subfigure}
\caption{The computational element $\Omega_k$ with solution points in black, flux points in red, and mapping points in blue, for a $\mathcal{P}2$ discretization.}
\label{fig:Elements}
\end{figure}

In this study, the solution and flux points are located at tensor products of Gauss points for quadrilateral elements and Williams-Shunn points \cite{williams2014symmetric} for triangular elements, as depicted in Figure \ref{fig:Elements}. 
The solution is approximated at each solution point, and then, the solution polynomial within each element is interpolated using nodal basis functions, ensuring element-wise continuity of the solution,
\begin{equation}
\pmb{u}_k^\delta (\pmb{\xi},t) = \sum_{i=1}^{N_p} \pmb{u}_{k,i}^{\delta} \phi_i (\pmb{\xi}) ,
\end{equation}
where $\pmb{u}_{k,i}^{\delta}$ is the numerical solution at point $i$ within element $\Omega_k$, $N_p$ is the total number of solution points within the element $\Omega_k$, and $\phi_i (\pmb{\xi})$ is the nodal basis function at point $i$. Furthermore, the flux polynomial is interpolated using nodal basis functions
\begin{equation}
\pmb{F}_k^{\delta D} (\pmb{\xi},t) = \sum_{i=1}^{N_p} F_{k,i}^{\delta} \phi_i (\pmb{\xi}) ,
\end{equation}
where $F_{k,i}^{\delta} = f \left( U_{k,i}^{\delta} , \nabla U_{k,i}^{\delta} \right)$ is the numerical flux value at point $i$ within element $\Omega_k$. The constructed numerical flux function, $\pmb{F}_k^{\delta D} (\pmb{\xi},t)$, is allowed to be discontinuous across cell interfaces, and the superscript $D$ denotes this discontinuity. Thus, a common Riemann flux must be defined to replace the normal flux. In this study, a Rusanov/Lax-Friedrichs flux is used at the interface between elements. To account for the jumps across cells, we follow Wang's formulation \cite{wang2009unifying} for simplex elements. By defining a correction field, $\vartheta_k \in \mathbb{P}^{\mathcal{P}}$, Eq. (\ref{equation_conservation_law}) is re-written within each element and must be satisfied at each solution point, i.e.,
\begin{equation}
\frac{d \pmb{u}_{k,i}^\delta}{d t} + \left( \pmb{\nabla} \cdot \pmb{F}_{k}^\delta \right)_{\pmb{\xi}_{k,i}} + \vartheta_{k,i} = 0.
\label{EqConsDiff}
\end{equation}
The correction field ensures a globally continuous flux polynomial and can be determined for each solution point, $i$, within element $k$, by
\begin{equation}
\vartheta_{k,i} = \frac{1}{|\Omega_k|} \sum_{f \in \partial \Omega_k} \sum_{j} \alpha_{i,f,j} \left[ \tilde{\pmb{F}} \right]_{f,j} S_f,
\end{equation}
where $f$ denotes the faces of the element $\Omega_k$, $j$ is the index for flux points, $\alpha_{i,f,j}$ are constant lifting coefficients, $\left[ \tilde{\pmb{F}} \right]_{f,j}$ is the difference between a common Riemann flux at point $j$ and the value of the internal flux, and $S_{f}$ is the area of the face $f$. The lifting coefficients are computed using a weighting function, $W$, and are independent of both geometry and the solution \cite{wang2009unifying}. In this study, the DG method is recovered via the FR formulation by choosing nodal basis functions as the weighting function \cite{wang2009unifying}, and the Rusanov and second method of Bassi and Rebay (BR2) are used for the common inviscid and viscous flux.

\subsection{Mesh Adaptive Direct Search Optimization}

In this study, the MADS optimization technique is used, which falls between the Generalized Pattern Search (GPS) \cite{torczon1997convergence} and the Coope and Price frame-based methods \cite{coope2001convergence}. Unlike GPS, MADS allows for a more flexible exploration of the design space during the optimization process, which makes it a more effective solution for both unconstrained and linearly constrained optimization \cite{audet2006mesh}. A major advantage of MADS over GPS is the flexible local exploration, known as poll directions, rather than a fixed set of directions. Two parameters are defined in the context of the MADS optimization: the mesh size parameter, $\Delta^m$, and the poll size parameter, $\Delta^p$.  The mesh size parameter determines the resolution of the design space mesh. A higher resolution leads to a more precise search while a lower resolution allows for a wider search and a higher chance of finding the global optimal solution.  The poll size parameter determines the neighborhood size around the incumbent point for selecting new trial points. The number of trial points per design cycle can be either $n+1$, known as minimal positive basis, or $2n$, known as maximal positive basis \cite{audet2006mesh}, where $n$ is the number of design variables. In this study, the minimal positive basis construction is used. 

The MADS algorithm consists of two sequential steps in each iteration: the search step and the poll step. Initially, the optimization procedure starts with the search step at the initial design point, $\pmb{\mathcal{X}}_0 = [\mathcal{X}^1_0, \mathcal{X}^2_0, ..., \mathcal{X}^n_0]$, where the subscript is the optimization iteration and the superscript denotes each design parameter. Pseudo-random trial points are generated, and infeasible ones, which are points within the design space not meeting the optimization problem's constraints, are discarded. The trial points are generated based on the current mesh and the direction vectors, $d_j \in \mathcal{D}$ (for $j = 1,2,...,n$), where $\mathcal{D}$ is the design space. $\mathcal{D}$ must be a positive spanning set \cite{davis1954theory}, and each direction, $d_j$, must be the product of some fixed non-singular generating matrix by an integer vector \cite{audet2006mesh}. The mesh at iteration $k$ is defined as \cite{audet2006mesh}
\begin{equation}
\mathcal{M}_k = \bigcup_{\mathcal{X} \in \mathcal{S}_k} \left\lbrace \mathcal{X}  + \Delta^m_k \mathcal{D} z : z \in \mathbb{N}^{n_{\mathcal{D}}} \right\rbrace ,
\end{equation}
where $\mathcal{S}_k$ is the set of trial points that the objective function is evaluated at, in iteration $k$. The mesh $\mathcal{M}_k$ is constructed from a finite set of $n_{\mathcal{D}}$ directions, $\mathcal{D} \subset \mathbb{R}^n$, scaled by a mesh size parameter $\Delta^m_k \in \mathbb{R}_{+}$. The objective function is evaluated at these trial points.
The iteration terminates either after evaluating the objective function at all trial points or upon finding a lower objective function, where the latter is employed in this study. Then, the next iteration starts with a new incumbent solution $\pmb{\mathcal{X}}_{k+1} \in \Omega$ with objective function of $\mathcal{F}(\pmb{\mathcal{X}}_{k+1}) < \mathcal{F}(\pmb{\mathcal{X}}_k)$, and a mesh size parameter $\Delta^m_{k+1} \geq \Delta_k^m$. The maximum value of the mesh size parameter, at any iteration, is set to one, $\Delta^m_{max}=1$. Note that the design space of each design variable is scaled to one, and a mesh size parameter of one can cover the entire design space.

On the other hand, if the search step fails to find a new optimum, the poll step is invoked before terminating the current optimization iteration. In the poll step, the mesh size parameter is reduced to define a new set of trial points closer to the incumbent design. The key difference between GPS and MADS is the new poll size parameter, $\Delta^p_k \in \mathbb{R}_{+}$, that controls the magnitude of the distance between trial points generated by the poll step to the incumbent point. This new set of trial points defined in the poll step is called a frame. The MADS frame at iteration $k$ is defined to be \cite{audet2006mesh}
\begin{equation}
P_k = \left\lbrace \mathcal{X}_k + \Delta^m_k d: d \in \mathcal{D}_k \right\rbrace \subset \mathcal{M}_k ,
\end{equation}
where $\mathcal{D}_k$ is a positive spanning set. In each MADS iteration, the mesh and poll size parameters are defined. The mesh size parameter of the new iteration is defined as \cite{audet2006mesh}
\begin{equation}
\Delta_{k+1}^m = 
\begin{cases}
\frac{1}{4} \Delta_k^m & \text{if the search step fails to find an improved design point,} \\
4 \Delta_k^m & \text{if an improved design point is found, and if } \Delta_k^m \leq \frac{1}{4}, \\
\Delta_k^m & \text{otherwise.} 
\end{cases}
\end{equation}
These rules ensure $\Delta_k^m$ is always a power of $4$ and never exceeds $1$. The poll size parameter is also defined as \cite{audet2006mesh}
\begin{equation}
\Delta_{k+1}^p = 
\begin{cases}
n \sqrt{\Delta_k^m} & \text{if the minimal positive basis construction is used,} \\
\sqrt{\Delta_k^m} & \text{if the maximal positive basis construction is used.} 
\end{cases}
\label{eq:poll_and_mesh_parameters}
\end{equation}

\begin{figure}
\centering
\begin{subfigure}{\textwidth}
\includegraphics[width=\textwidth]{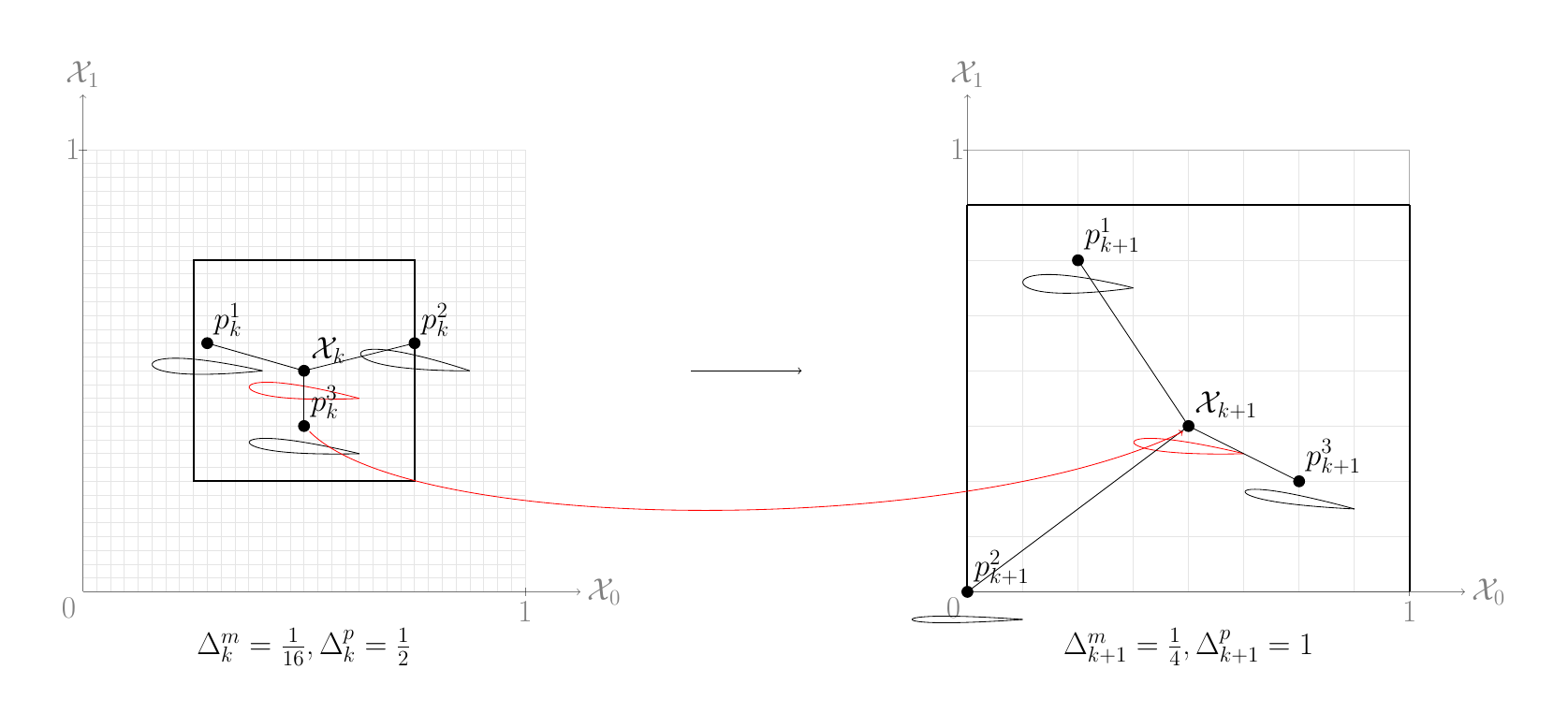}
\subcaption{Search step.}
\label{figure_mads_search}
\end{subfigure}
\begin{subfigure}{\textwidth}
\includegraphics[width=\textwidth]{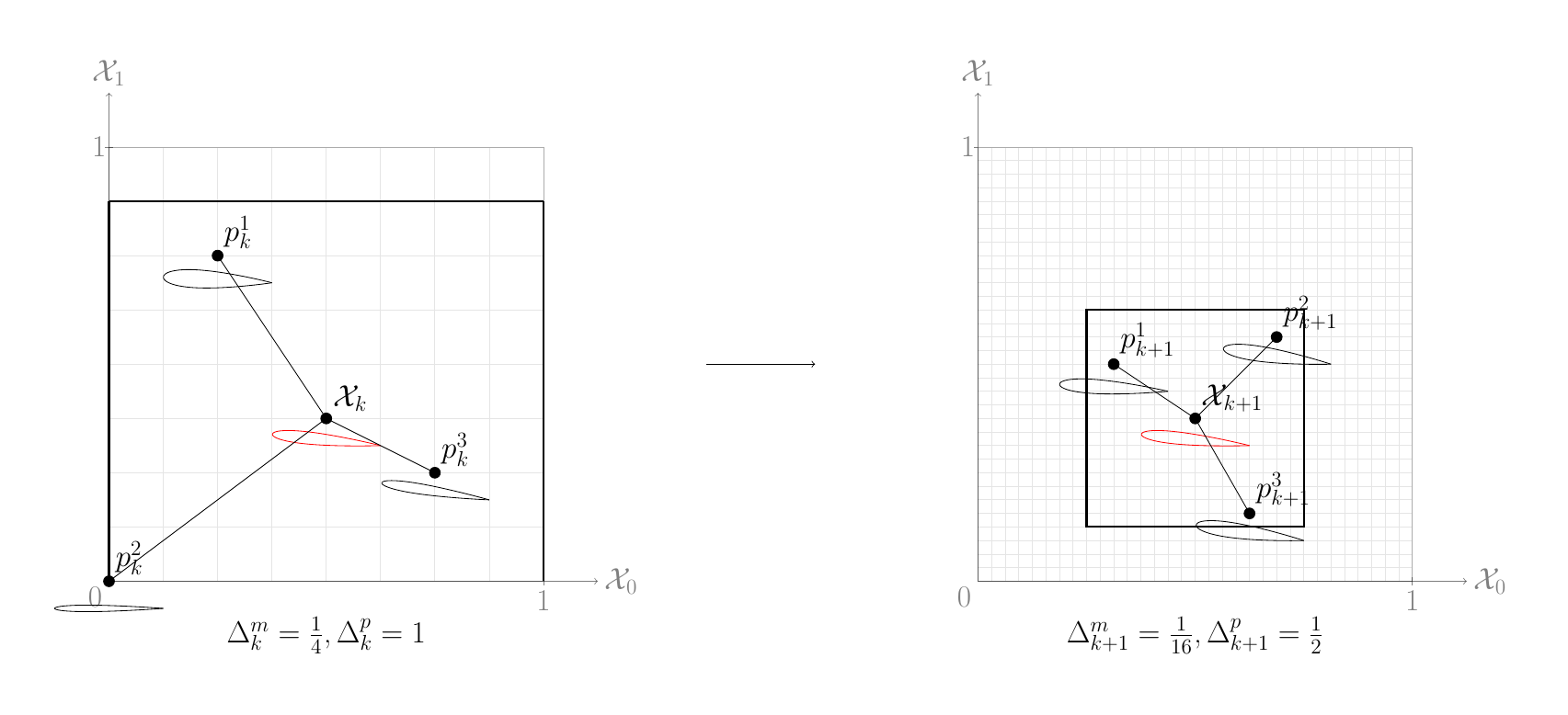}
\subcaption{Poll step.}
\label{figure_mads_poll}
\end{subfigure}
\caption{Search and poll steps of the MADS optimization techniques for iteration $k$ for the NACA example.}
\label{figure_mads_steps}
\end{figure}

As an example, which will be considered in detail in the numerical results section, consider a NACA 4-digit airfoil where the design parameters are the airfoil's thickness $\mathcal{X}_1 \in [6, 20]$ and the angle of attack $\mathcal{X}_2 \in [0^\circ,12^\circ]$. In this case, the bottom left corner in the design space $[\mathcal{X}_1,\mathcal{X}_2] = [0,0]$ shown in Figure \ref{figure_mads_steps} corresponds to the minimum thickness and angle of attack. In contrast, the top right corner $[\mathcal{X}_1,\mathcal{X}_2] = [1,1]$ would correspond to the maximum thickness and angle of attack. Our objective is to find the point in the allowable design space that produces minimal noise. Starting from some initial design $\mathcal{X}_0$ and initial mesh and poll size parameters, we start by searching in the poll region at three candidate designs $p_k^1, p_k^2, p_k^3$. If, for example, $p_k^3$ is a configuration that produces less noise, we re-center our search around this point as our new optimal design, and increase the mesh size parameter for the subsequent step, as shown in Figure \ref{figure_mads_search}. However, if none of these points are quieter, we maintain the current optimal design, and shrink the mesh and poll size parameters for the subsequent step. This means that designs closer to the current optimum, i.e. more similar thickness and angle of attack, will be searched in the next step.
Finally, when a new incumbent objective function is found in a design iteration, the optimization may converge or continue depending on the stopping criterion. The optimization problem is terminated when the stopping criteria are met.

\section{Deep Cavity}
\label{sec:DeepCavity}

\begin{figure}
\centering
\includegraphics[width=0.7\textwidth]{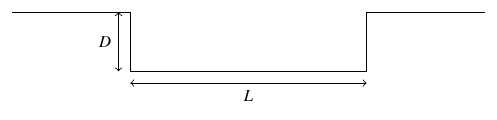}
\caption{The geometry of a two-dimensional cavity.}
\label{fig:CavityGeometry}
\end{figure}

Flow over a two-dimensional deep cavity is a classical problem in fluid mechanics and aeroacoustics, and has been the subject of extensive research due to its relevance for a range of engineering applications. The flow over a cavity is characterized by a complex interplay between the boundary layer, the recirculation zone inside the cavity, and the external flow. The presence of the cavity can lead to a variety of aerodynamic and aeroacoustic phenomena, such as flow separation, unsteady vortex shedding, and acoustic resonance. Understanding the aerodynamic and aeroacoustic characteristics of flow over a cavity is crucial for optimizing the design and performance of many engineering systems. This topic has been studied using various techniques, including experiments \cite{gharib1987effect}, CFD simulations \cite{colonius1999numerical, rowley2002self}, and aeroacoustic analysis \cite{larsson2004aeroacoustic, ask2009sound, martin2019noise}. The geometry of a cavity is typically given in terms of the length-to-depth ratio, $L/D$, depicted in Figure \ref{fig:CavityGeometry}. The Reynolds number is usually based on the depth of the cavity, $Re_D = U_\infty D / \nu$, where $U_\infty$ is the free-stream velocity and $\nu$ is the kinematic viscosity. The numerical simulation is first validated using the numerical reference study \cite{larsson2004aeroacoustic}, and then the optimization procedure is explained.

\subsection{Validation}

In this section, the flow simulation over an open deep cavity is validated by comparing the time history of drag coefficient, pressure perturbation coefficient, and the overall sound pressure level at observer locations with the reference data \cite{larsson2004aeroacoustic}.  

\subsubsection{Computational Details}

\begin{figure}
\centering
\begin{subfigure}{0.7\textwidth}
\includegraphics[width=\textwidth]{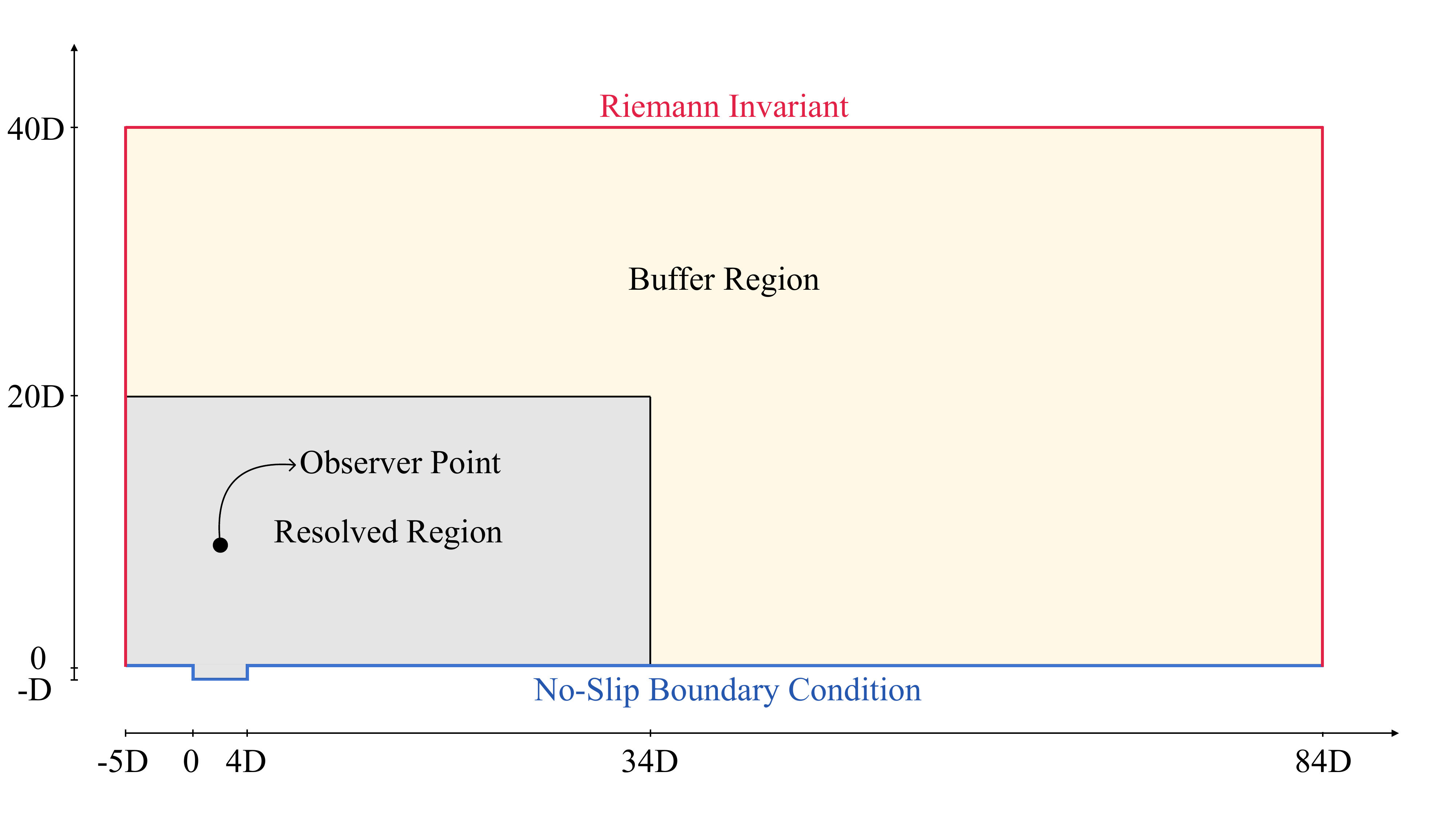}
\subcaption{The geometry.}
\end{subfigure}
\begin{subfigure}{0.7\textwidth}
\includegraphics[width=\textwidth]{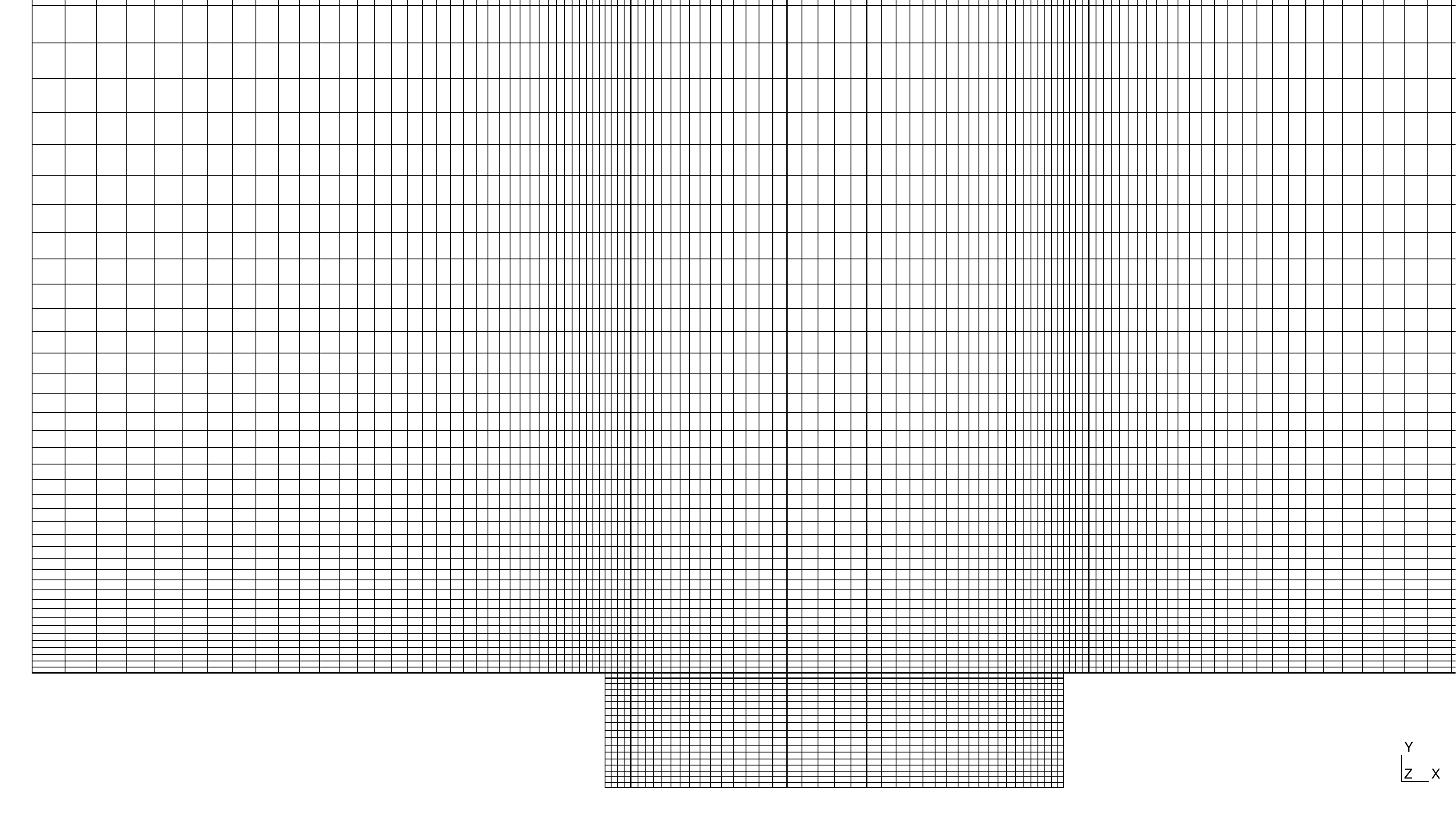}
\subcaption{The mesh around the cavity.}
\end{subfigure}
\caption{The computational domain of the open cavity.}
\label{fig:CavityGeo}
\end{figure}

The entrance length of the domain is set to $5D$, which affects the oscillation regime significantly and results in a shear layer mode with the time-averaged boundary layer thickness of $\delta \approx 0.3D$ at the cavity entrance. The outflow boundary is $80D$ away from the downstream cavity wall, where the last $50D$ of the downstream domain acts as a buffer region to eliminate the reflections of the acoustic waves from the computational boundaries. The resolved domain in the $y$-direction extends between $0 < y/D < 20$, and the buffer region extends between $20 < y/D < 40$. Stretching ratios of $1.05$ and $1.075$ are used in the resolved and buffer regions, respectively, with the smallest element size of $0.05D$ inside the cavity. A total of $13,076$ quadrangular elements are used to validate the open cavity with $\mathcal{P}2$, resulting in $117,684$ solution points. The boundary conditions of the domain, along with the cavity's geometry and mesh, are shown in Figure \ref{fig:CavityGeo}. The length-to-depth ratio of the cavity is $L/D=4$, the Reynolds number based on the cavity depth is $Re_D = 1500$, and the inflow Mach number is $M_\infty=0.15$. 

\subsubsection{Results and Discussion}

\begin{figure}
\centering
\includegraphics[width=0.7\textwidth]{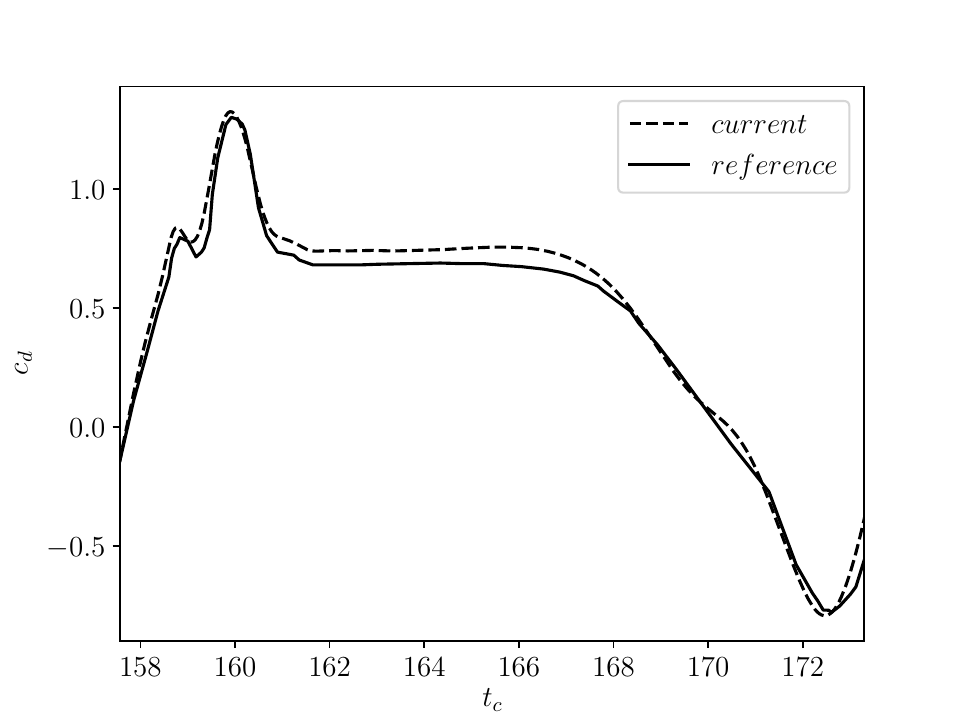}
\caption{The time history of drag coefficient of the open cavity.}
\label{fig:Cavity2DCD}
\end{figure}

The simulation is run for $100 t_c$, where $t_c = D/U_\infty$ and $U_\infty$ is the free-stream velocity, to allow initial transients to disappear and the simulation to reach a fully developed behavior.  The drag coefficient of the open cavity is defined as
\begin{equation}
c_d = \frac{F_x}{\frac{1}{2} \rho_\infty U_\infty^2 D},
\end{equation}
where $\rho_\infty$ is the free-stream density and $F_x$ is force per unit width in the $x$-direction and is computed on the three cavity walls. The drag coefficient of the open cavity is plotted against the convective time in Figure \ref{fig:Cavity2DCD},  which is in good agreement with the reference \cite{larsson2004aeroacoustic}.

\begin{figure}
\centering
\includegraphics[width=0.7\textwidth]{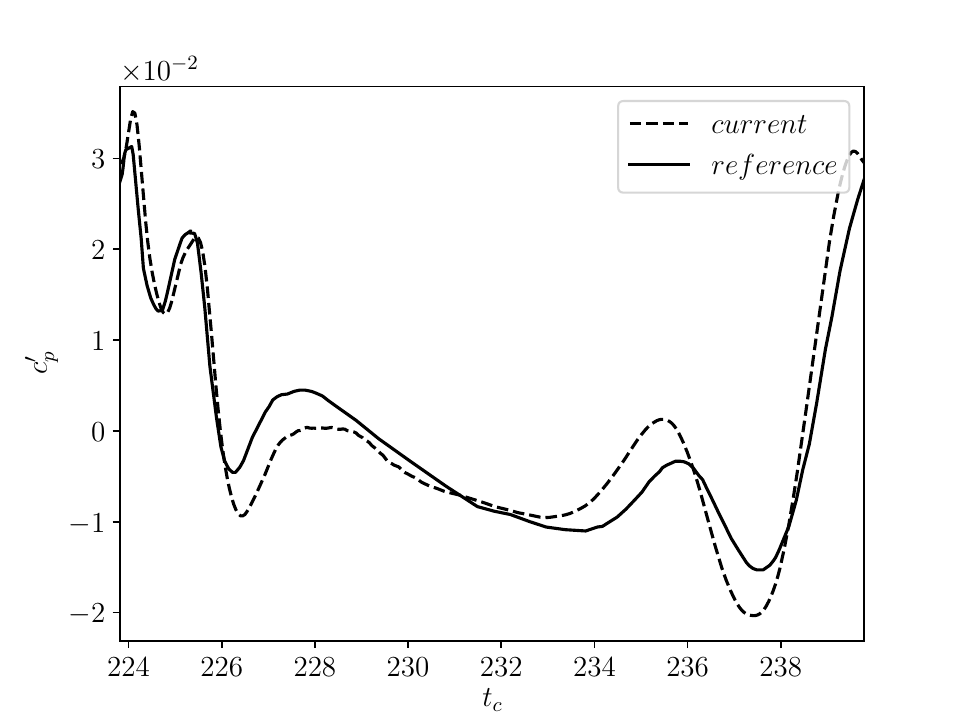}
\caption{The pressure perturbation coefficient of the open cavity at $[x/D, y/D] = [1, 7.16]$.}
\label{fig:Cavity2DCP}
\end{figure}

\begin{figure}
\centering
\includegraphics[width=0.7\textwidth]{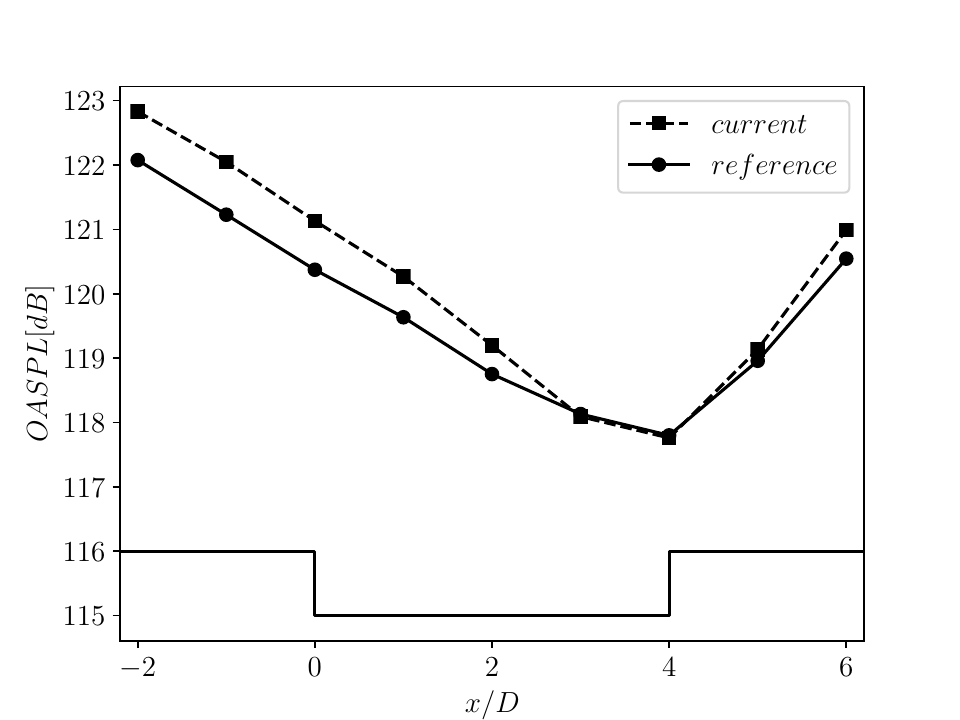}
\caption{The OASPL of the open cavity.}
\label{fig:Cavity2DSPL}
\end{figure}

The pressure perturbation coefficient is defined as
\begin{equation}
c_p^\prime = c_p - \overline{c_p},
\end{equation}
where $\overline{c_p}$ is the time-averaged pressure coefficient, and $c_p$ is the instantaneous pressure coefficient defined as
\begin{equation}
c_p = \frac{p - p_\infty}{\frac{1}{2} \rho_\infty U_\infty^2},
\end{equation}
where $p$ is the static pressure, and $p_\infty$ is the free-stream pressure. The $c^\prime_p$ is plotted against $t_c$ for one convective time at an observer point located at $[x/D=1$, $y/D=7.16]$ in Figure \ref{fig:Cavity2DCP}. And, finally, the OASPL at a set of observer points is computed and shown in Figure \ref{fig:Cavity2DSPL},  where excellent agreement with the reference results is observed near the trailing edge of the cavity. However, some minor discrepancies are observed at data points further upstream, which are also apparent in the time series of the pressure perturbation coefficient provided in Figure \ref{fig:Cavity2DCP}.  As noted in this prior work, the acoustics of open cavities is highly-sensitive to the incoming boundary layer. Best efforts were made to match the mean velocity profiles provided in \cite{larsson2004aeroacoustic}, but the available reference configuration is not precise enough to assess whether an exact match is obtained.

The periodicity of the flow can be described by the fundamental or Strouhal frequency, where $St = 2.444$ based on the length of the cavity, is in excellent agreement with the reported value of $2.45$ \cite{ask2009sound}.

\begin{figure}
\centering

\begin{subfigure}{\textwidth}
\centering
\begin{subfigure}{0.45\textwidth}
\includegraphics[width=\textwidth]{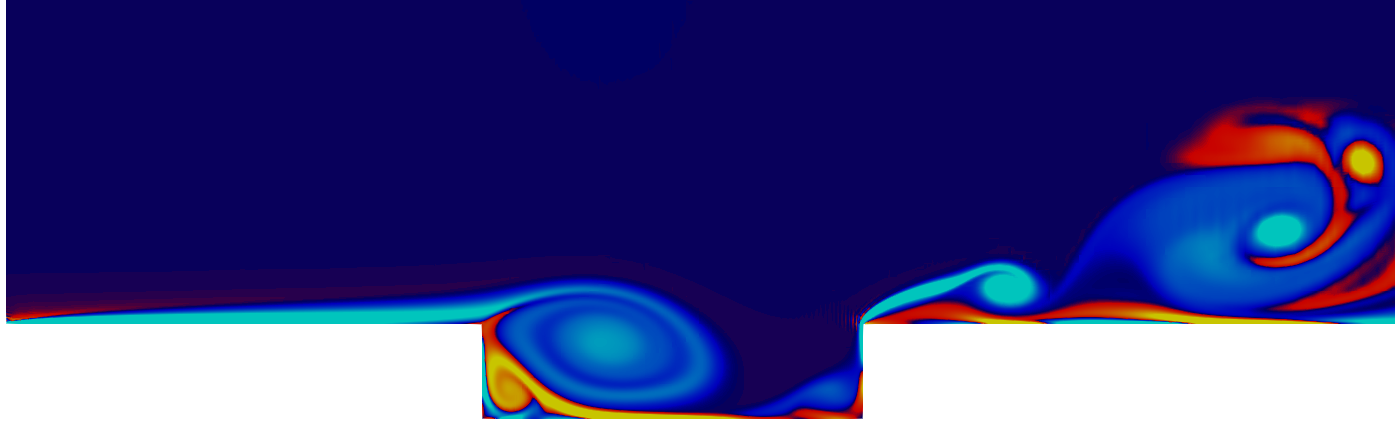}
\end{subfigure}
\begin{subfigure}{0.45\textwidth}
\includegraphics[width=\textwidth]{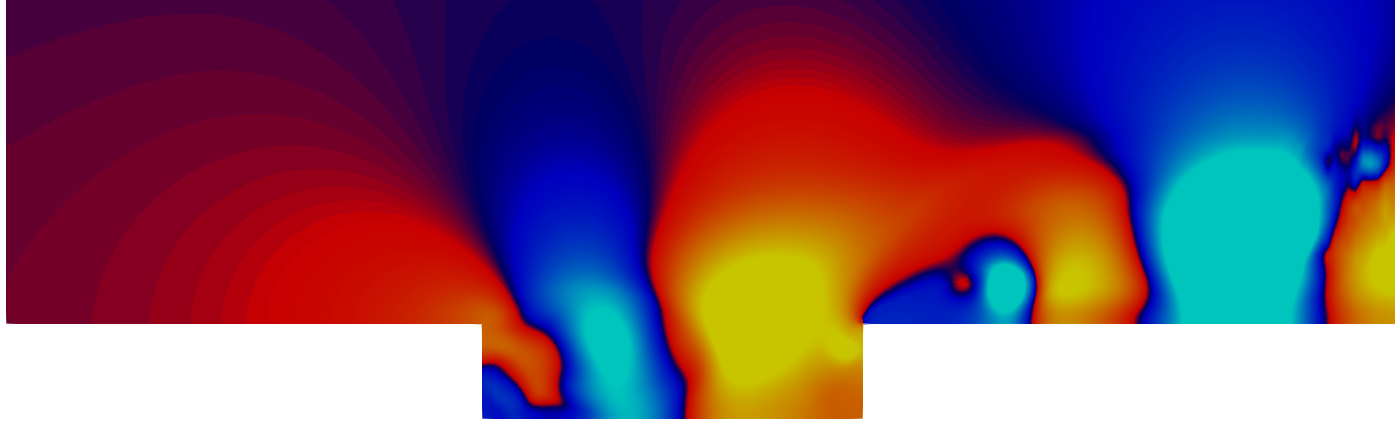}
\end{subfigure}
\subcaption{$t_c=100$.}
\label{fig:Cavity2DVorticity_a}
\end{subfigure}

\begin{subfigure}{\textwidth}
\centering
\begin{subfigure}{0.45\textwidth}
\includegraphics[width=\textwidth]{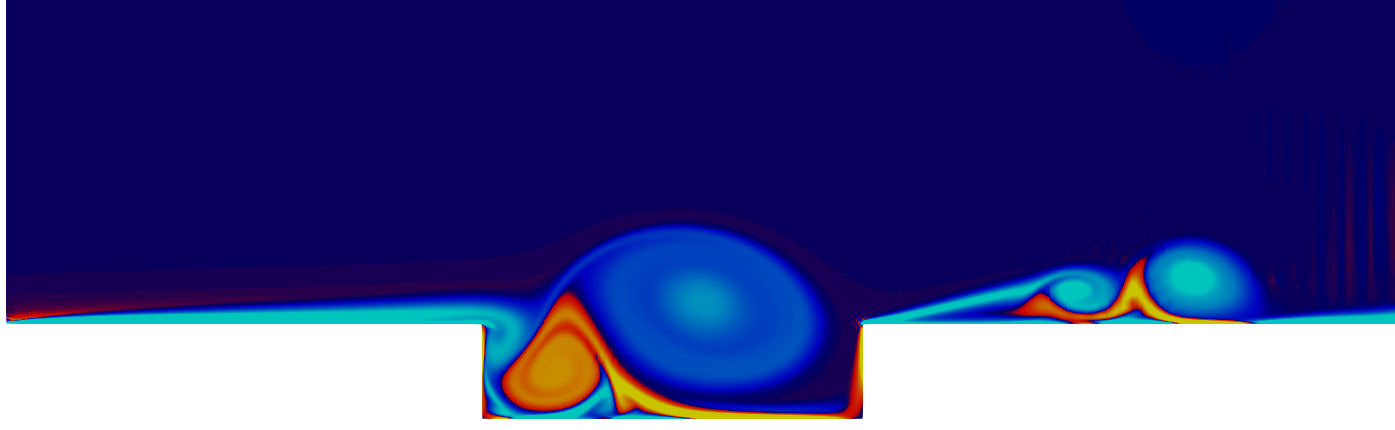}
\end{subfigure}
\begin{subfigure}{0.45\textwidth}
\includegraphics[width=\textwidth]{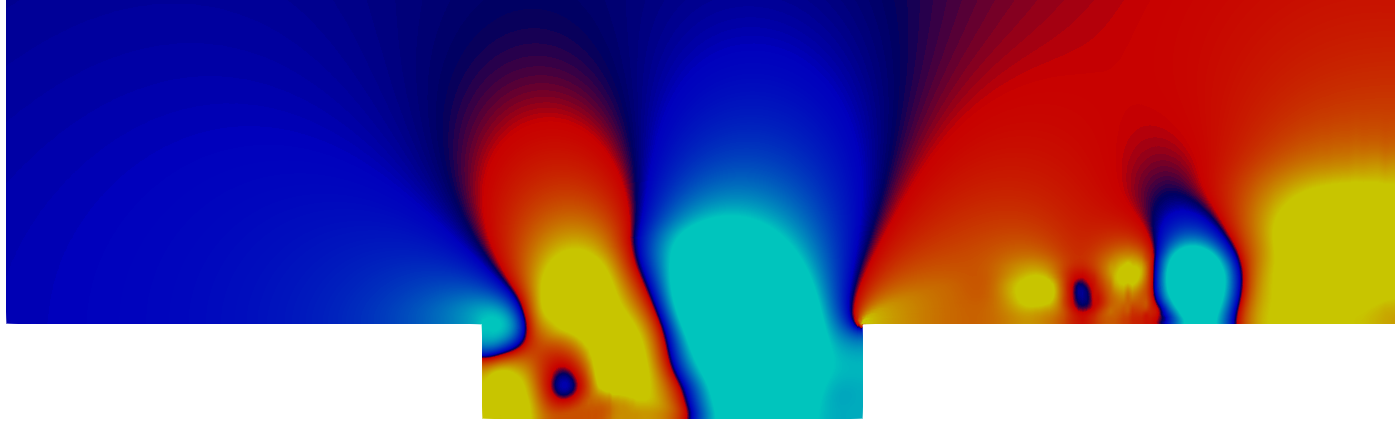}
\end{subfigure}
\subcaption{$t_c=106.2$.}
\label{fig:Cavity2DVorticity_b}
\end{subfigure}

\begin{subfigure}{\textwidth}
\centering
\begin{subfigure}{0.45\textwidth}
\includegraphics[width=\textwidth]{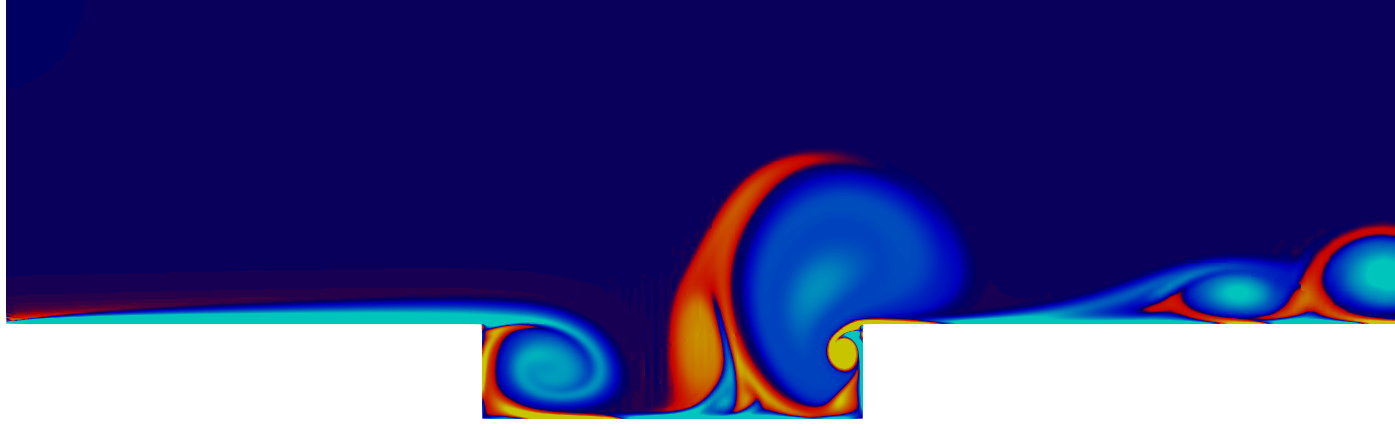}
\end{subfigure}
\begin{subfigure}{0.45\textwidth}
\includegraphics[width=\textwidth]{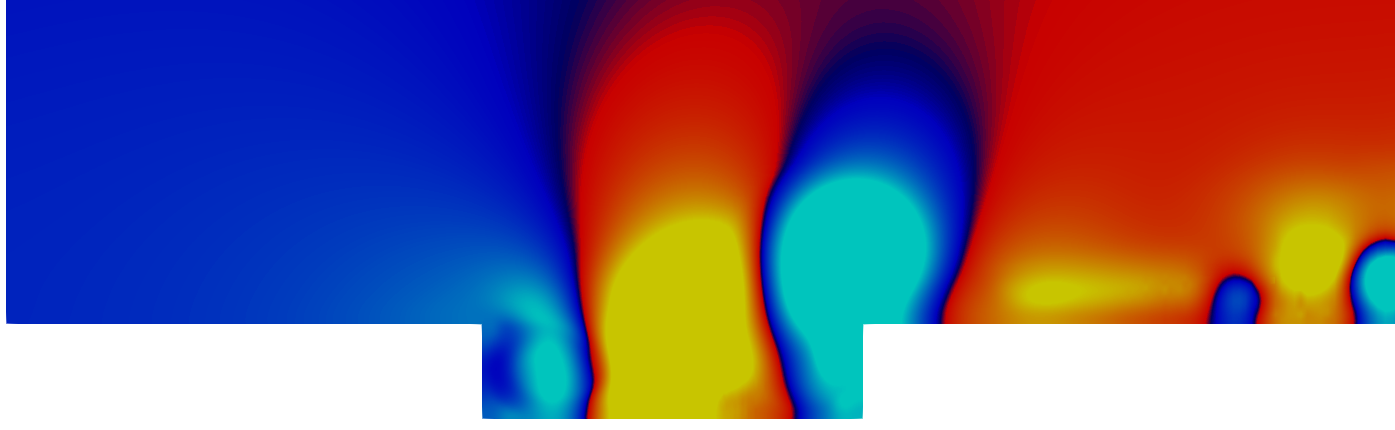}
\end{subfigure}
\subcaption{$t_c=109.7$.}
\label{fig:Cavity2DVorticity_c}
\end{subfigure}

\begin{subfigure}{\textwidth}
\centering
\begin{subfigure}{0.45\textwidth}
\includegraphics[width=\textwidth]{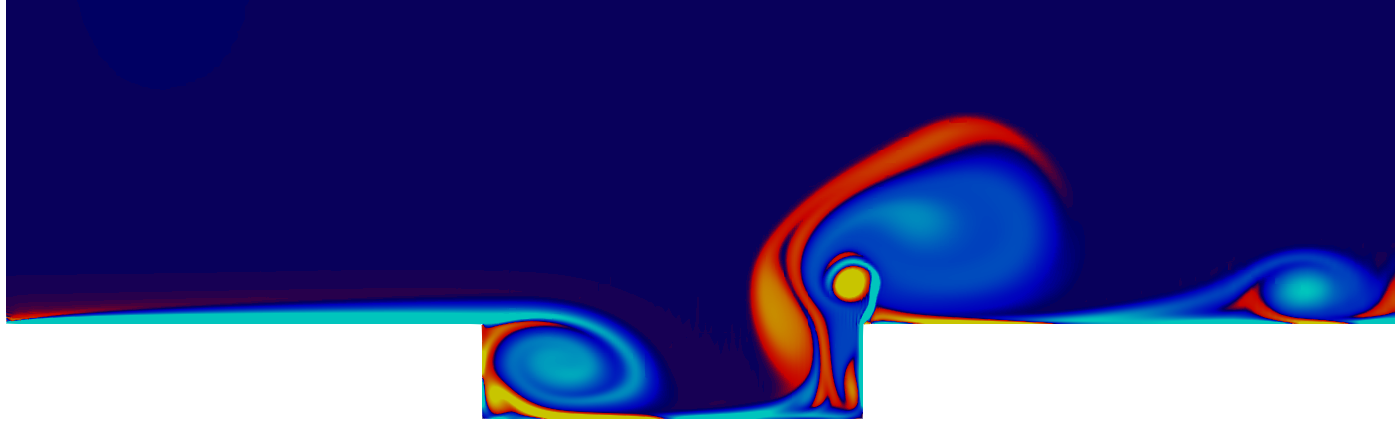}
\end{subfigure}
\begin{subfigure}{0.45\textwidth}
\includegraphics[width=\textwidth]{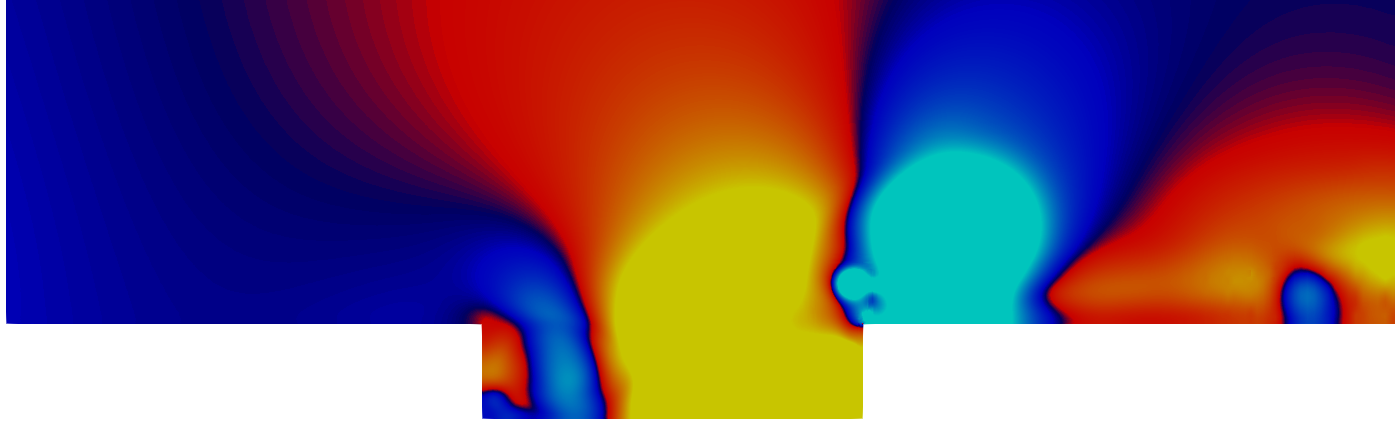}
\end{subfigure}
\subcaption{$t_c=111.4$.}
\label{fig:Cavity2DVorticity_d}
\end{subfigure}

\begin{subfigure}{\textwidth}
\centering
\begin{subfigure}{0.45\textwidth}
\includegraphics[width=\textwidth]{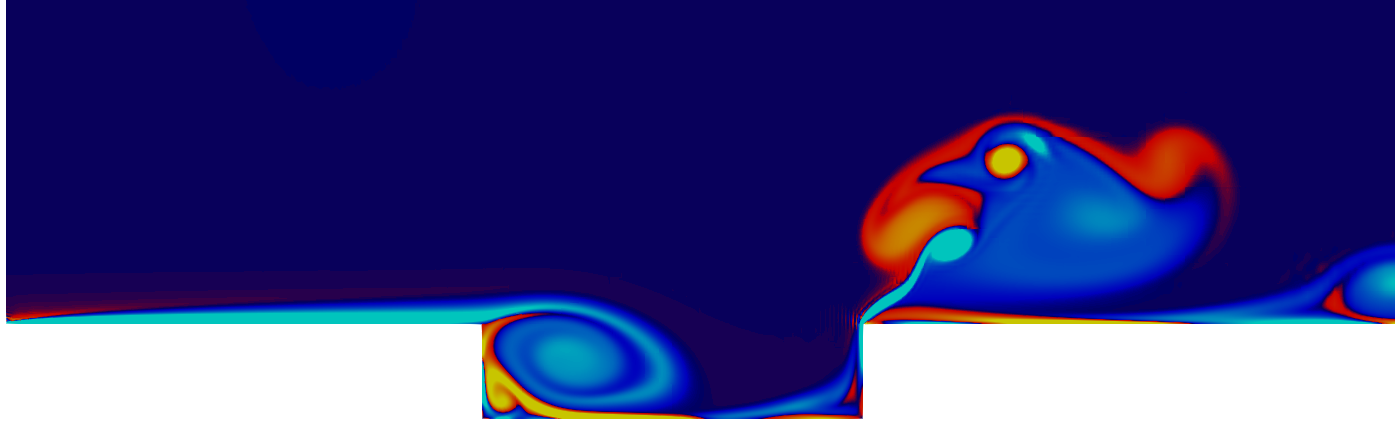}
\end{subfigure}
\begin{subfigure}{0.45\textwidth}
\includegraphics[width=\textwidth]{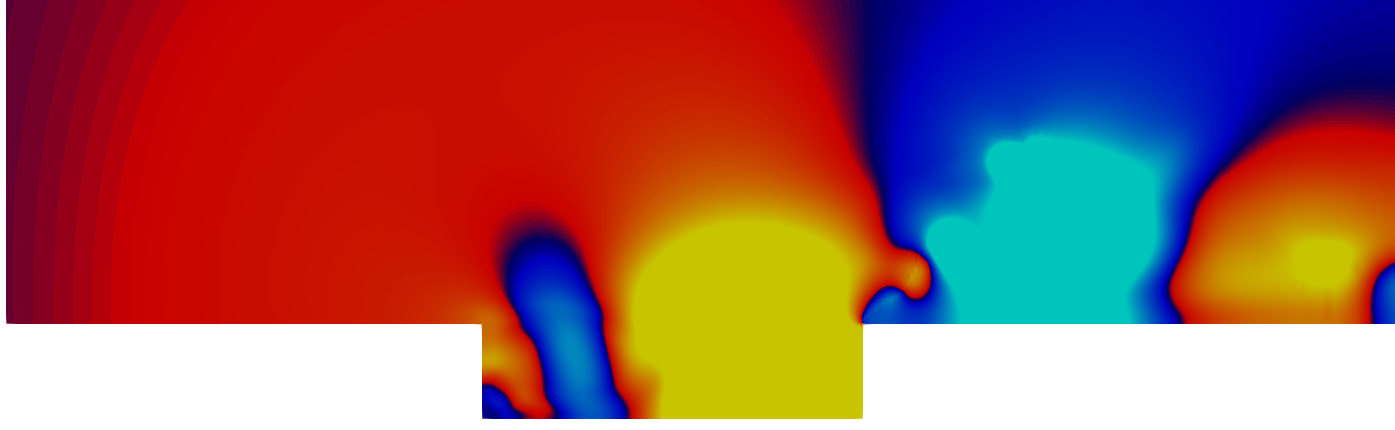}
\end{subfigure}
\subcaption{$t_c=113.2$.}
\label{fig:Cavity2DVorticity_e}
\end{subfigure}

\begin{subfigure}{\textwidth}
\centering
\begin{subfigure}{0.45\textwidth}
\centering
\includegraphics[width=\textwidth]{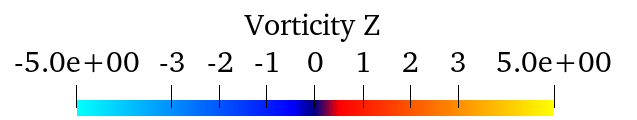}
\end{subfigure}
\begin{subfigure}{0.45\textwidth}
\centering
\includegraphics[width=\textwidth]{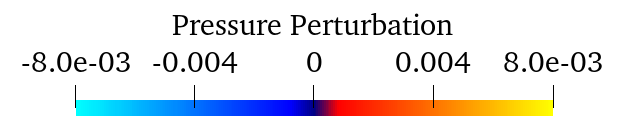}
\end{subfigure}
\end{subfigure}

\caption{Z-component of the vorticity and pressure perturbation snapshots of the open cavity.}
\label{fig:Cavity2DVorticity}
\end{figure}

Vortex structures and flow patterns are shown in Figure \ref{fig:Cavity2DVorticity} at different times. The first and dominant vortex is shed from the upstream cavity inlet and evolves from the recirculation bubble at the cavity inlet, shown in Figure \ref{fig:Cavity2DVorticity_a}. This vortex grows rapidly within the cavity while connected to the cavity's leading edge. Growth ceases when the vortex leaves the leading edge, and the free stream is injected into the cavity and upstream of the vortex (Figure \ref{fig:Cavity2DVorticity_b}). As the primary clockwise-rotating vortex hits the trailing edge, a new counter-clockwise-rotating vortex is shed into the cavity at the trailing edge (Figure \ref{fig:Cavity2DVorticity_c}), producing a high amplitude acoustic wave. The new counter-clockwise-rotating vortex then moves downstream (Figure \ref{fig:Cavity2DVorticity_d}), cutting the primary vortex (Figure \ref{fig:Cavity2DVorticity_e}). The primary vortex dictates the cavity flow's fundamental frequency and plays a crucial role in the sound generation of such flows \cite{larsson2004aeroacoustic}.

\begin{figure}
\centering
\includegraphics[width=0.7\textwidth]{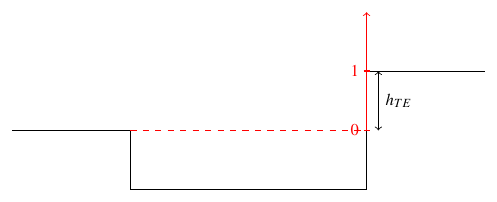}
\caption{The height of the two-dimensional cavity's trailing edge, used as a design variable.}
\label{fig:CavityGeometry_hTE}
\end{figure}

In the next section, the height of the cavity trailing edge wall, shown in Figure \ref{fig:CavityGeometry_hTE}, is optimized to reduce the sound perceived by an observer located at $[x/D, y/D] = [2, 7.16]$. 

\subsection{Optimization}

In this section, the noise at the observer point located at $\pmb{x}_{obs}/D = [2, 7.16]$ is minimized by changing the height of the cavity trailing edge wall, $h_{TE}$, using the MADS technique. Thus, $\pmb{\mathcal{X}}= \left[ h_{TE} \right]$ is the design variable and $\mathcal{X}_0=0$, while the objective function is the root-mean-squared of the pressure perturbation, $\mathcal{F} = p^\prime_{rms}$.

\subsubsection{Results and Discussion}

\begin{figure}
\centering
\begin{subfigure}{0.7\textwidth}
\includegraphics[width=\textwidth]{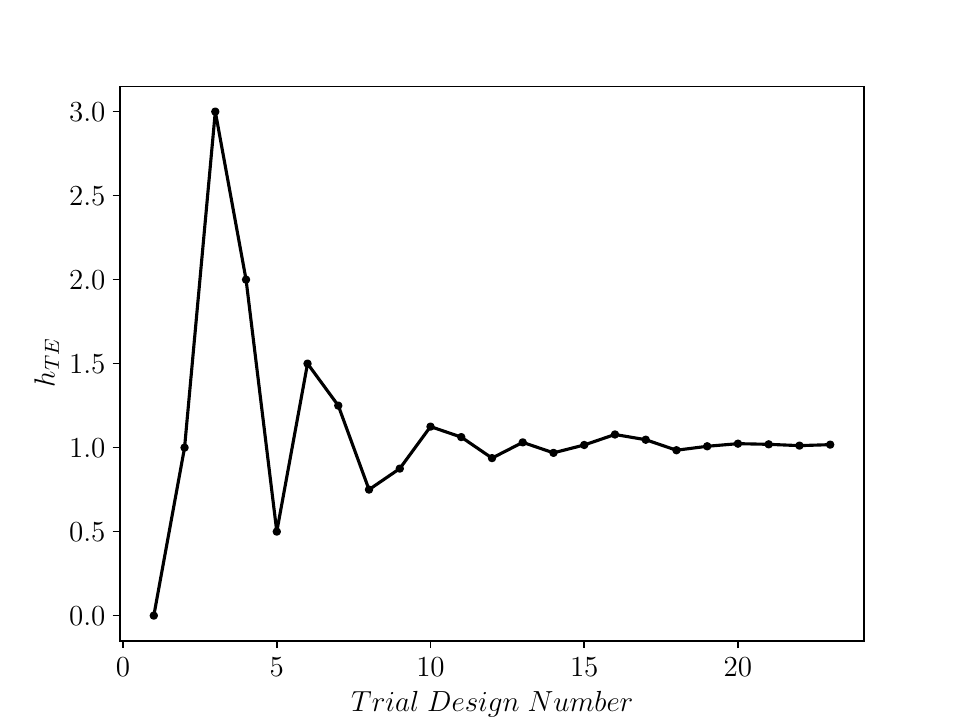}
\subcaption{The design space.}
\label{figure_cavity_optimization_design_space}
\end{subfigure}
\begin{subfigure}{0.7\textwidth}
\includegraphics[width=\textwidth]{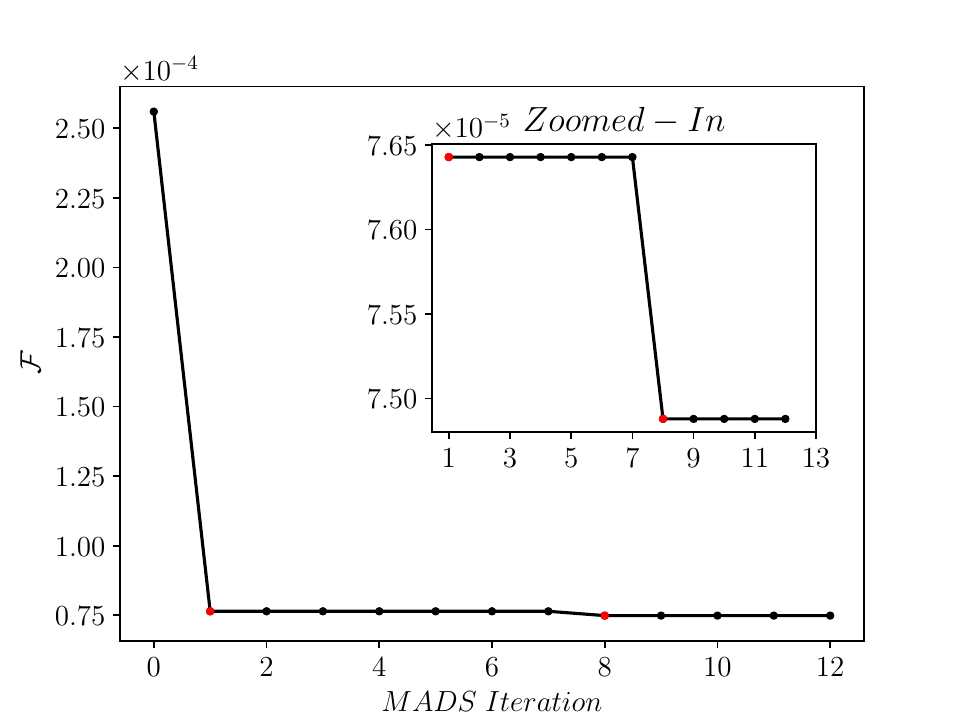}
\subcaption{The objective function convergence with the new incumbent designs highlighted in red.}
\label{figure_cavity_optimization_obj}
\end{subfigure}
\caption{The design space and objective function convergence for the open deep cavity. }
\label{figure_cavity_optimization}
\end{figure}

\begin{figure}
\centering
\includegraphics[width=0.7\textwidth]{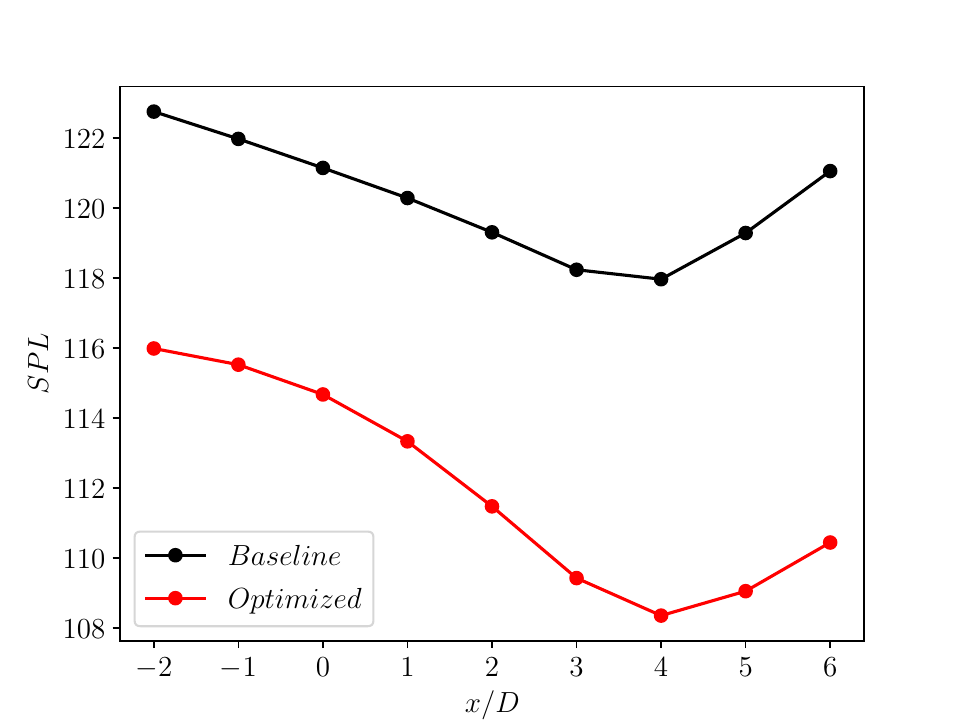}
\caption{The OASPL at different observer points for the optimum design of the open deep cavity.}
\label{fig:Cavity2DSPLOptimized}
\end{figure}

\begin{figure}
\centering

\begin{subfigure}{\textwidth}
\centering
\begin{subfigure}{0.45\textwidth}
\includegraphics[width=\textwidth]{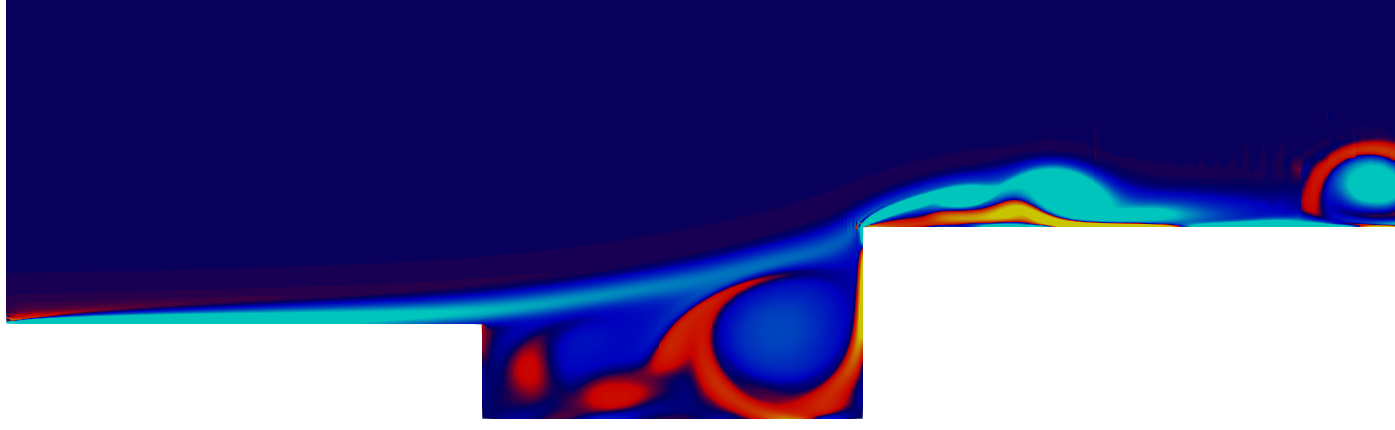}
\end{subfigure}
\begin{subfigure}{0.45\textwidth}
\includegraphics[width=\textwidth]{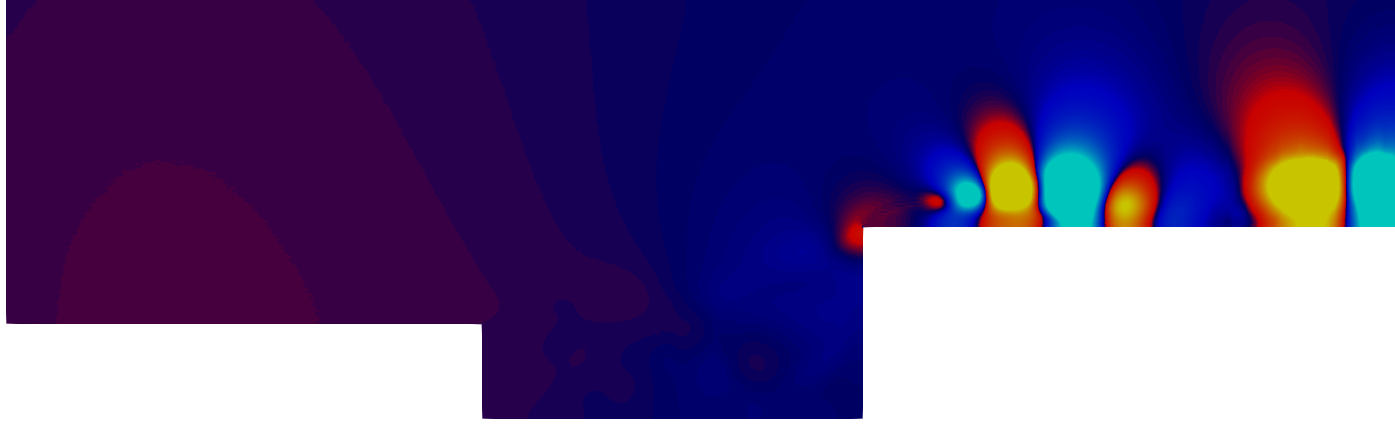}
\end{subfigure}
\subcaption{$t_c=101.3$}
\label{fig:Cavity2Doptimum_a}
\end{subfigure}

\begin{subfigure}{\textwidth}
\centering
\begin{subfigure}{0.45\textwidth}
\includegraphics[width=\textwidth]{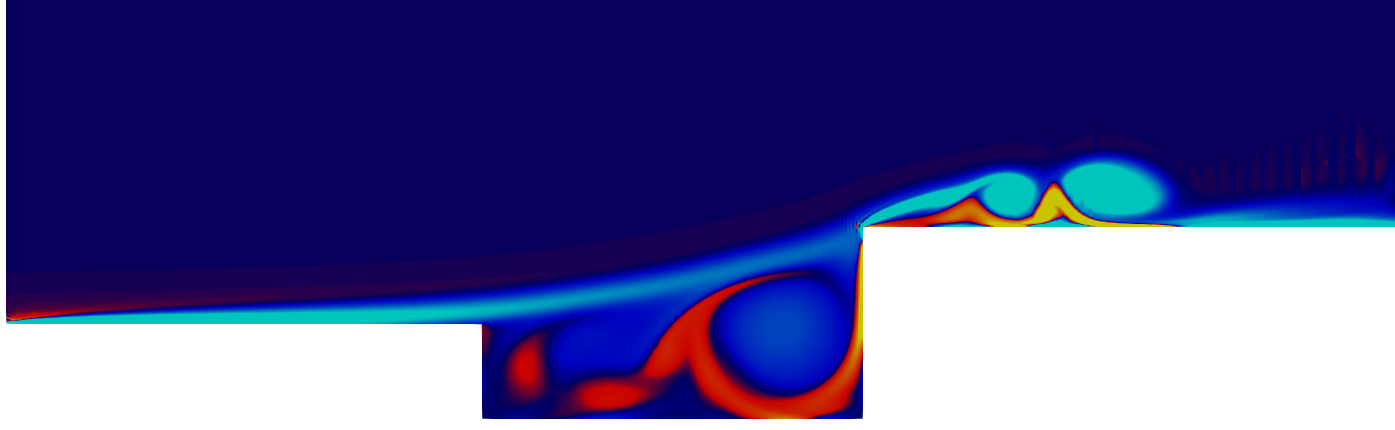}
\end{subfigure}
\begin{subfigure}{0.45\textwidth}
\includegraphics[width=\textwidth]{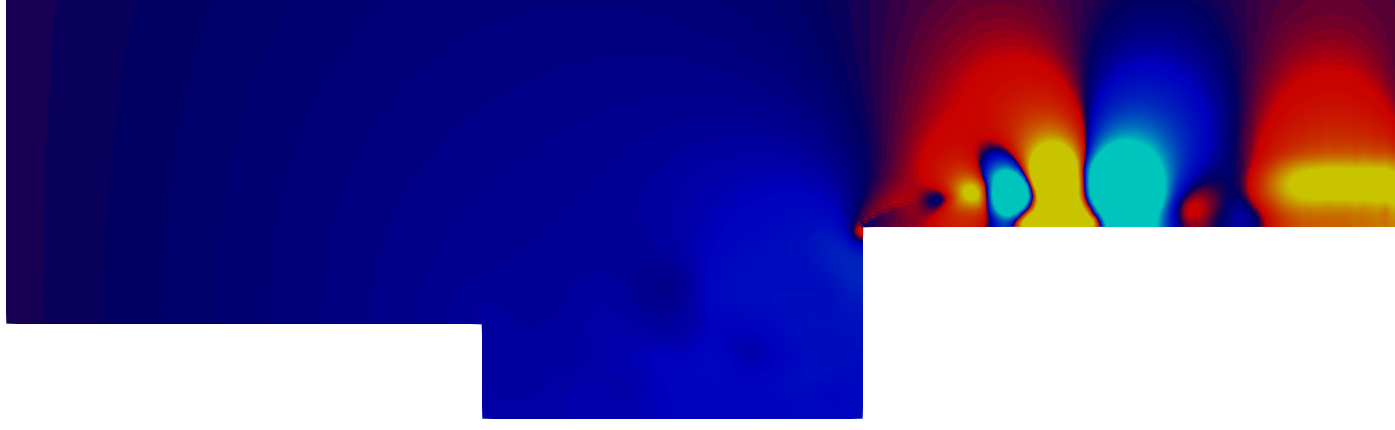}
\end{subfigure}
\subcaption{$t_c=102.6$.}
\label{fig:Cavity2Doptimum_b}
\end{subfigure}

\begin{subfigure}{\textwidth}
\centering
\begin{subfigure}{0.45\textwidth}
\includegraphics[width=\textwidth]{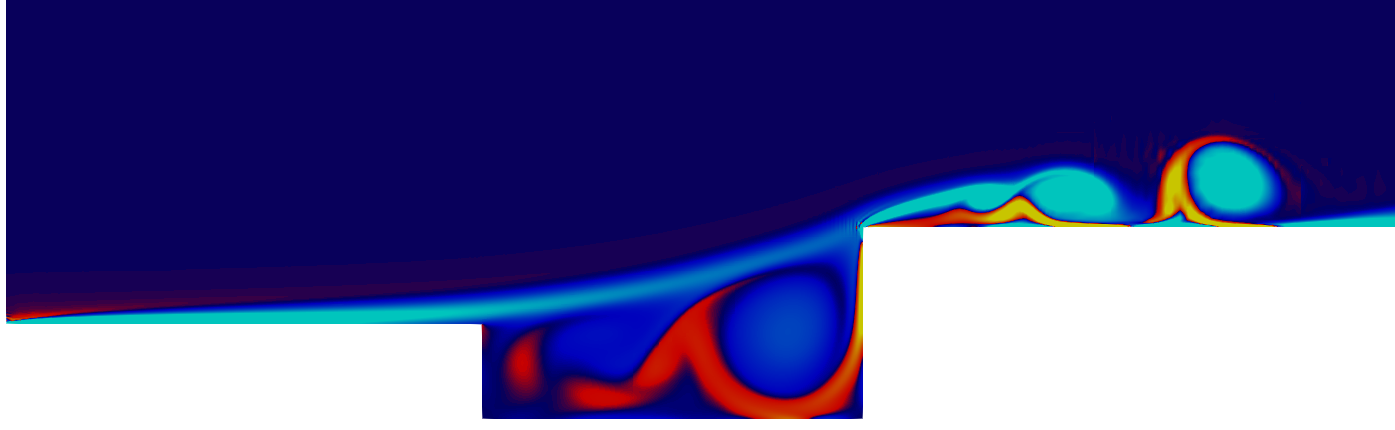}
\end{subfigure}
\begin{subfigure}{0.45\textwidth}
\includegraphics[width=\textwidth]{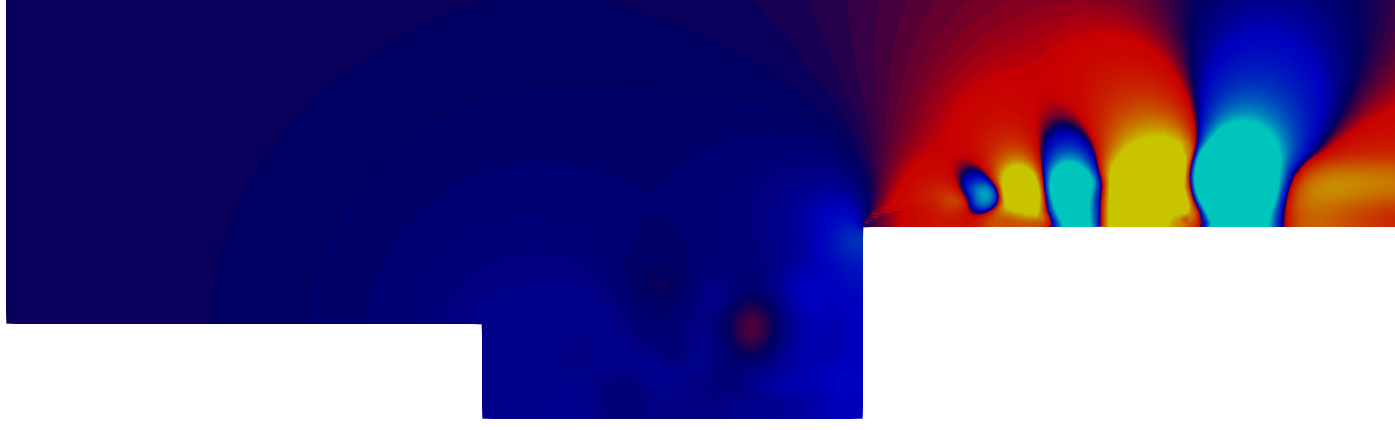}
\end{subfigure}
\subcaption{$t_c=105.1$.}
\label{fig:Cavity2Doptimum_c}
\end{subfigure}

\begin{subfigure}{\textwidth}
\centering
\begin{subfigure}{0.45\textwidth}
\includegraphics[width=\textwidth]{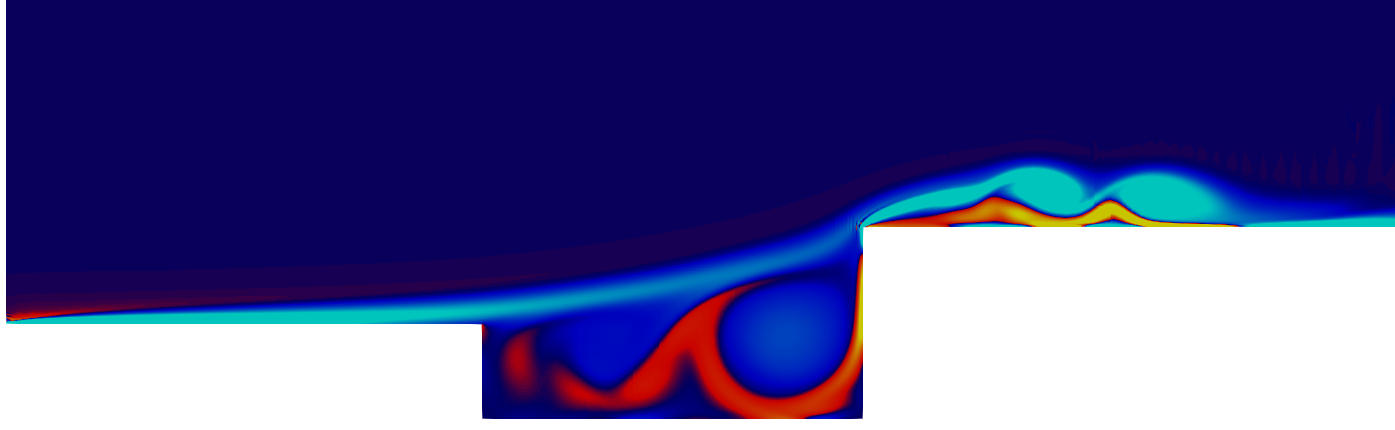}
\end{subfigure}
\begin{subfigure}{0.45\textwidth}
\includegraphics[width=\textwidth]{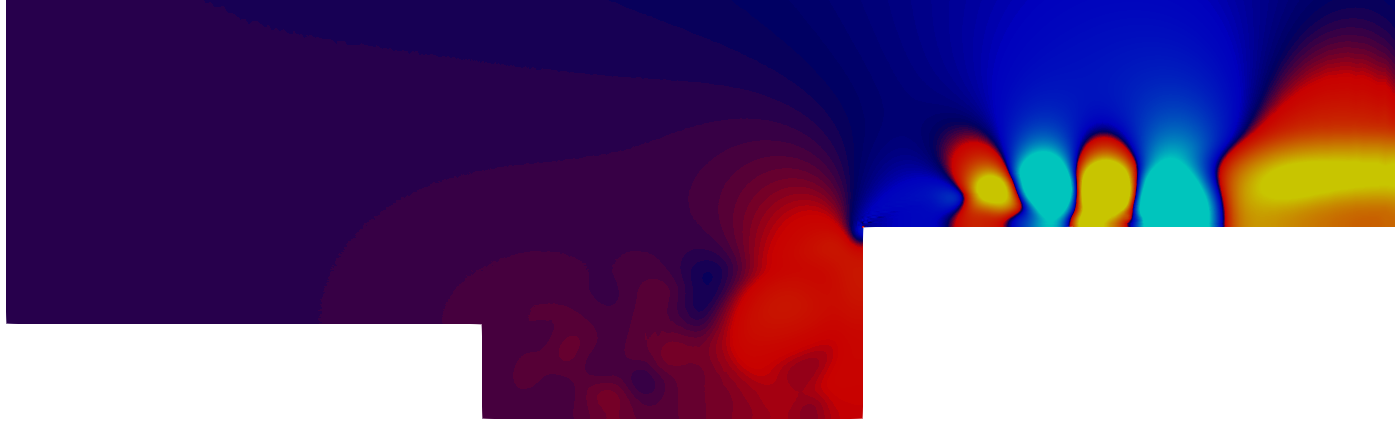}
\end{subfigure}
\subcaption{$t_c=108.8$.}
\label{fig:Cavity2Doptimum_d}
\end{subfigure}

\begin{subfigure}{\textwidth}
\centering
\begin{subfigure}{0.45\textwidth}
\includegraphics[width=\textwidth]{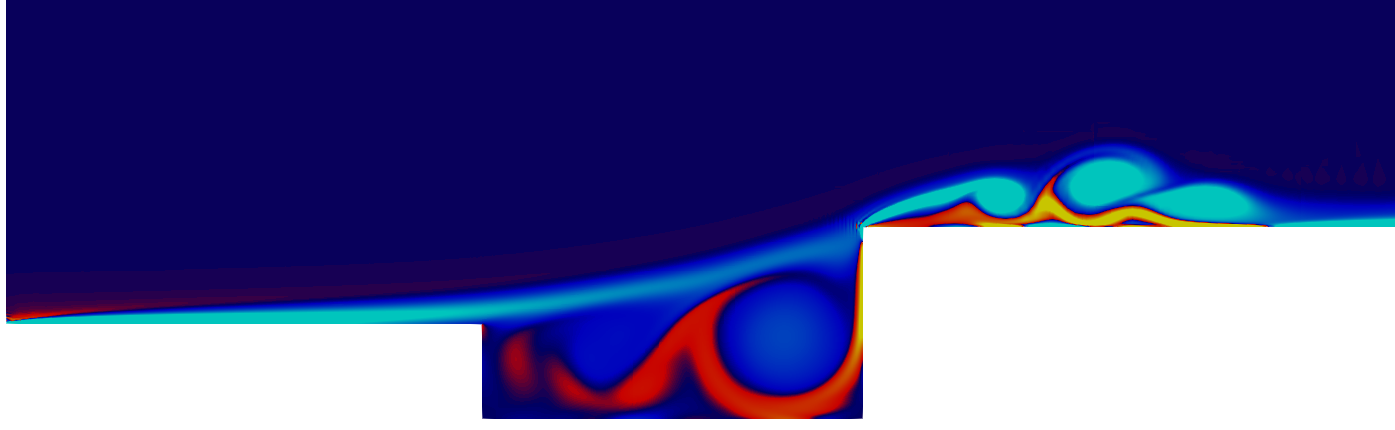}
\end{subfigure}
\begin{subfigure}{0.45\textwidth}
\includegraphics[width=\textwidth]{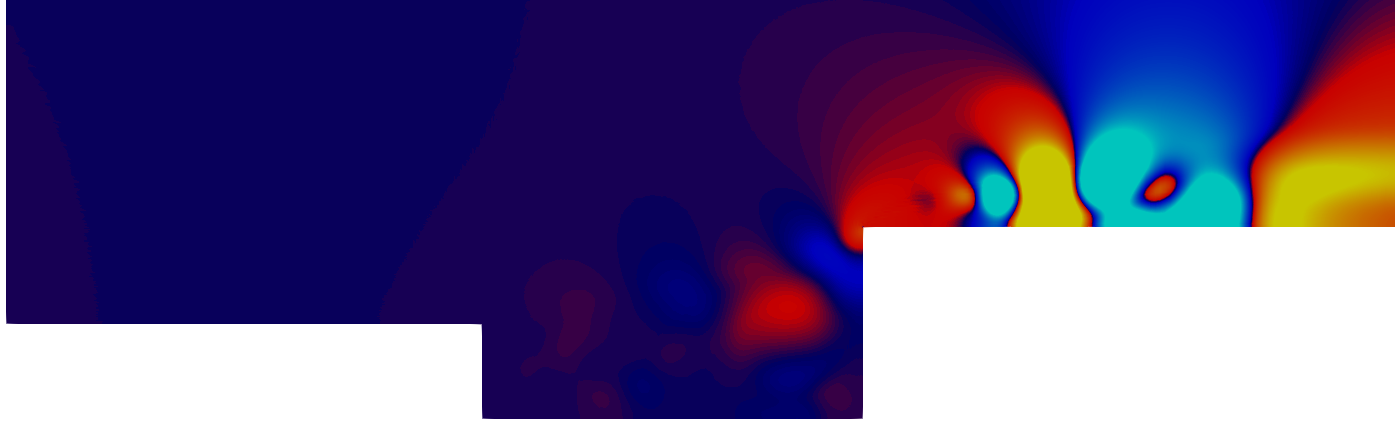}
\end{subfigure}
\subcaption{$t_c=110.3$.}
\label{fig:Cavity2Doptimum_e}
\end{subfigure}

\begin{subfigure}{\textwidth}
\centering
\begin{subfigure}{0.45\textwidth}
\centering
\includegraphics[width=\textwidth]{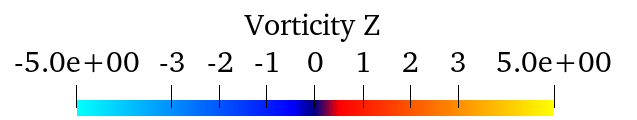}
\end{subfigure}
\begin{subfigure}{0.45\textwidth}
\centering
\includegraphics[width=\textwidth]{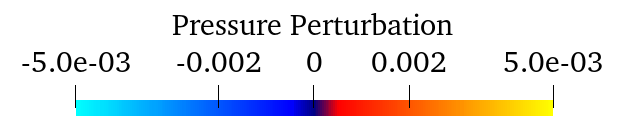}
\end{subfigure}
\end{subfigure}

\caption{Z-component of the vorticity and pressure perturbation snapshots of the optimized open cavity.}
\label{fig:Cavity2Doptimum}
\end{figure}

Upper and lower bounds of $0$ and $4$, respectively, are chosen for the design variable, $h_{TE}$, with the objective function being the root-mean-squared of the pressure perturbation at $\pmb{x}_{obs}$. The design variable converged to $h_{TE} = 1.0156$ after $12$ MADS iterations with a total of $23$ objective function evaluations. The design space and the objective function convergence are shown in Figure \ref{figure_cavity_optimization}. It can be seen from Figure \ref{figure_cavity_optimization_design_space} that a wide design space is investigated by the MADS optimization technique, and only two incumbent values are found, as shown in Figure \ref{figure_cavity_optimization_obj}. The overall sound pressure level is decreased to $111.43~dB$ for the optimum design, down from $119.3~dB$ of the baseline design. Thus, $7.87 dB$ decrease in the OASPL at the observer is achieved. It can be seen from Figure \ref{fig:Cavity2DSPLOptimized} that the sound at all other observer points is also reduced, as expected.
The z-component of the vorticity and pressure perturbation are plotted in Figure \ref{fig:Cavity2Doptimum} at different times. For the optimized shape of the open cavity, the primary clockwise vortex trapped inside the cavity reduces the emitted noise. However, there are vortices shedding off the trailing edge of the cavity; but their acoustic waves are much smaller in amplitude compared to those of the baseline design.

\section{Tandem Cylinders}
\label{sec:TandemCylinders}

\begin{figure}
\centering
\includegraphics[width=0.7\textwidth]{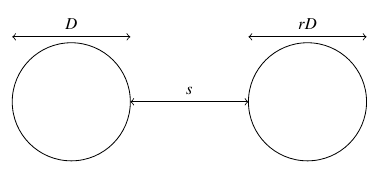}
\caption{The arrangement of the tandem cylinders.}
\label{fig:Cylinders2Dgeo}
\end{figure}

The flow around two tandem cylinders consists of multiple flow features including flow separation, reattachment, recirculation, and quasi-periodic vortex shedding, amongst others. The physics of such flows is highly dependent on the diameter ratio of the cylinders, the spacing between them, and the Reynolds number. The diameter ratio of the cylinders is defined as $r=D_{d}/D_{u}$, where $D_d$ and $D_u$ are the downstream and upstream diameter of the cylinders, respectively. The spacing of the cylinders, $s$, is defined as the distance between the rear of the upstream cylinder to the front of the downstream cylinder. These definitions are depicted in Figure \ref{fig:Cylinders2Dgeo}.

There is growing literature on flow over tandem cylinders \cite{shan2021effect, ding2007numerical, sharman2005numerical} and the resulting acoustic field \cite{fitzpatrick2003flow, mohany2009numerical, finnegan2010experimental}. In the next section, the flow features of two cylinders in a tandem configuration are investigated, along with the sound radiated by them, and compared with reference data \cite{ding2007numerical}. Finally, the diameter ratio between the cylinders and their distance are optimized to reduce the noise at the observer located $2D$ above the upstream cylinder.

\subsection{Validation}

In this section, the simulation of flow over tandem cylinders is validated by comparing the time history of lift and drag coefficients with the reference data \cite{ding2007numerical}. The relationship between the mean time-averaged drag coefficient of the cylinders with the space between them is investigated and compared to the available literature \cite{igarashi1981characteristics}.
 
\subsubsection{Computational Details}

\begin{figure}
\centering
\begin{subfigure}{0.45\textwidth}
\centering
\includegraphics[width=\textwidth]{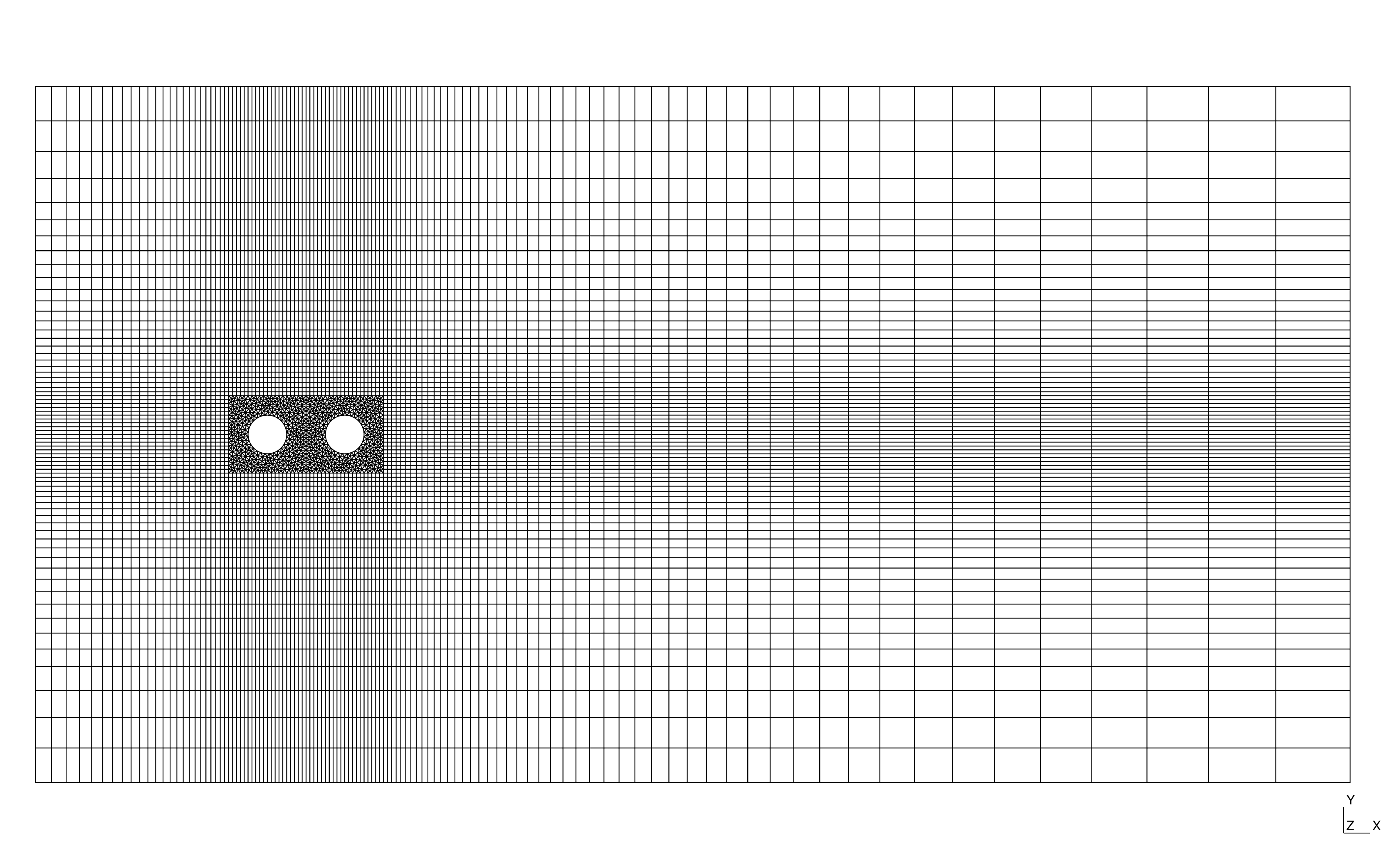}
\subcaption{The computational domain.}
\end{subfigure}
\begin{subfigure}{0.45\textwidth}
\centering
\includegraphics[width=\textwidth]{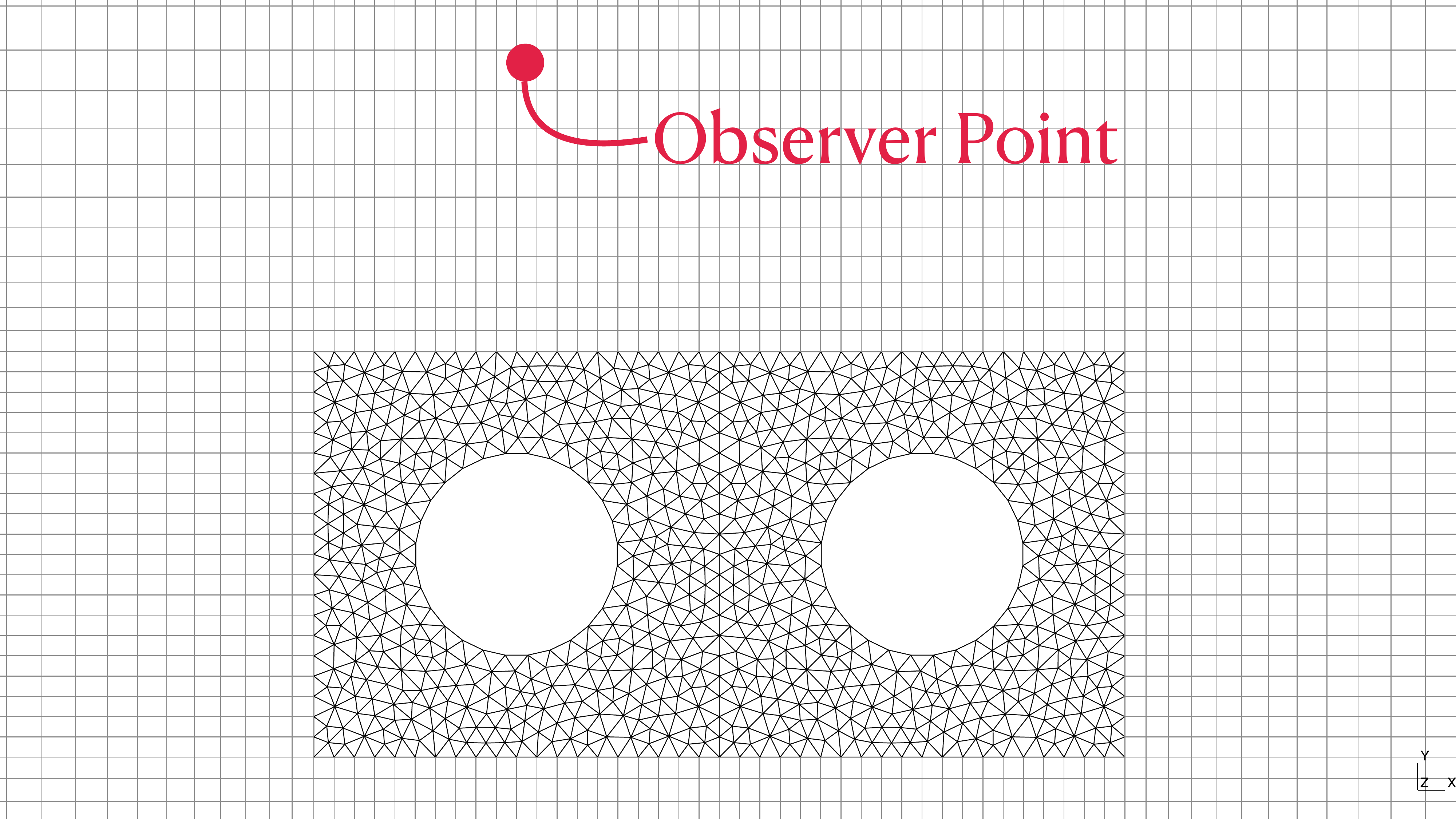}
\subcaption{The vicinity of the cylinders.}
\label{fig:CylindersObserver}
\end{subfigure}
\caption{The computational grid for tandem cylinders.}
\label{fig:Cylinders2Dmesh}
\end{figure}

The tandem cylinders configuration is first run for $r=1$ and $s=4.5$. A total of $8718$ triangular and quadrilateral elements are used, and the computational grid is shown in Figure \ref{fig:Cylinders2Dmesh}. The simulation is started with a $\mathcal{P}1$ simulation, switched to $\mathcal{P}2$ after $1500t_c$, and then, is run for $500t_c$ to compute the statistical characteristics of the flow. $t_c=D/U_\infty$ is the time needed for flow to pass the upstream cylinder or the convective time, and $U_\infty$ is the free-stream velocity. The Reynolds number for this study is $Re=200$, and the inflow Mach number is $M_\infty = 0.2$. The $6$-stage $5^{th}$-order accurate Explicit Singly Diagonally Implicit Runge-Kutta (ESDIRK) temporal scheme \cite{vermeire2017behaviour} is used to advance the simulation in time.

\subsubsection{Results and Discussion}

\begin{figure}
\centering
\begin{subfigure}{0.7\textwidth}
\includegraphics[width=\textwidth]{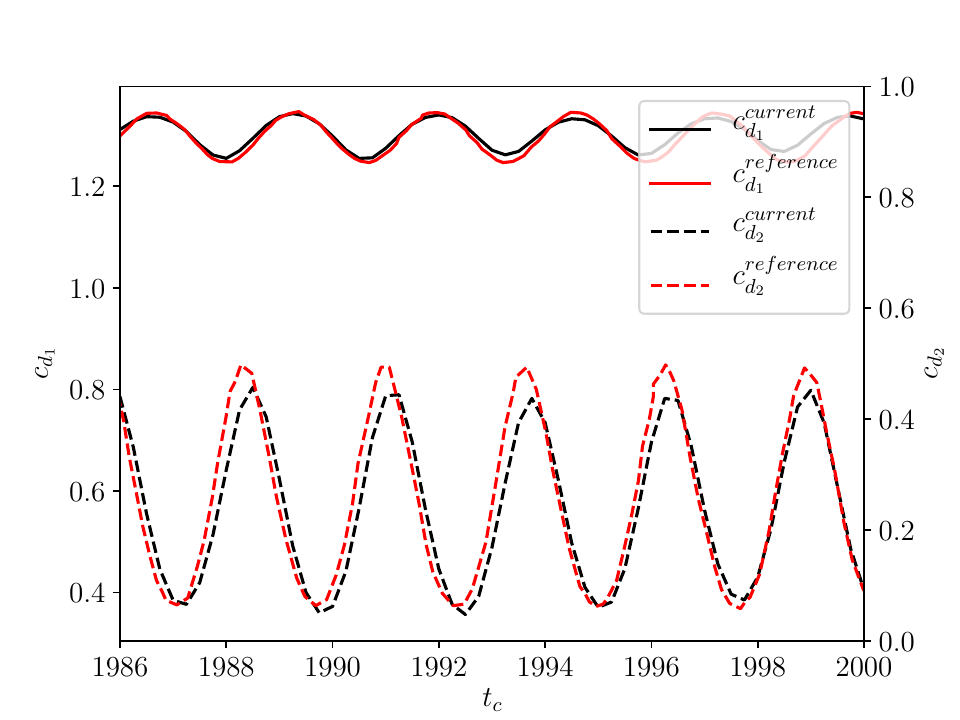}
\subcaption{The drag coefficient.}
\end{subfigure}
\begin{subfigure}{0.7\textwidth}
\includegraphics[width=\textwidth]{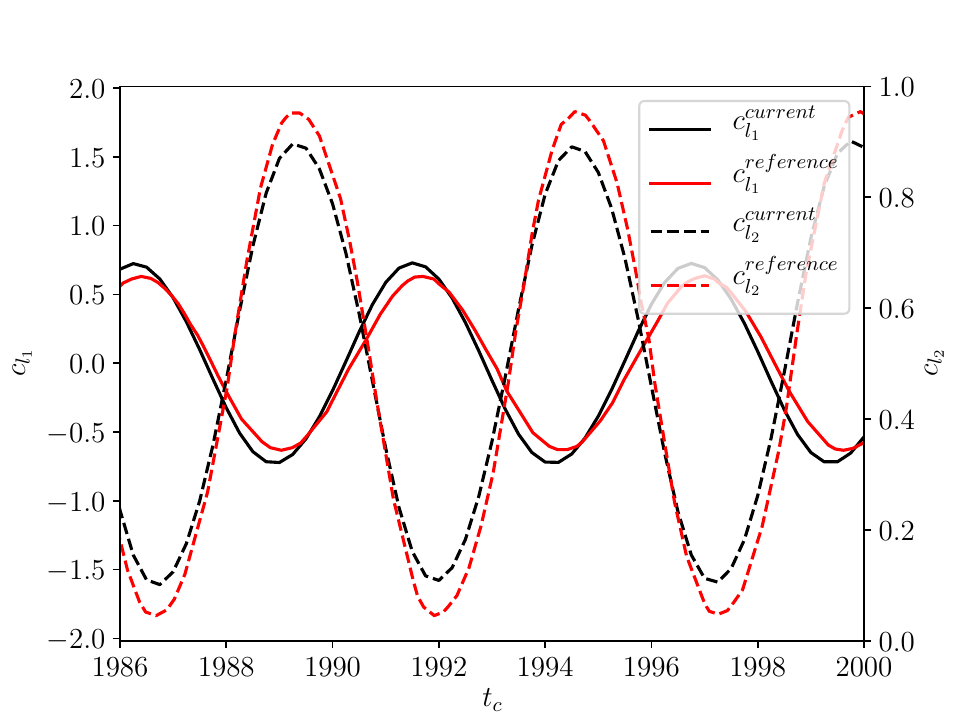}
\subcaption{The lift coefficient.}
\end{subfigure}
\caption{The lift and drag coefficients of flow past a pair of tandem cylinders $(s=4.5)$ at $Re=200$.}
\label{fig:Cylinders2Dvalidation}
\end{figure}

\begin{figure}
\centering
\includegraphics[width=0.7\textwidth]{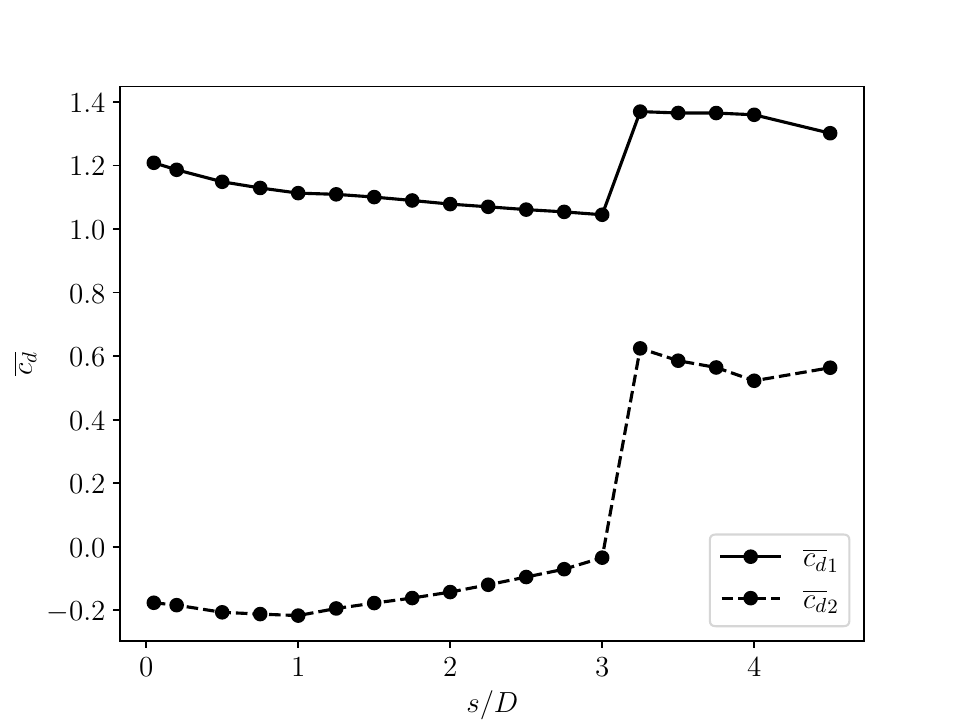}
\caption{Drag coefficient for the tandem cylinders versus separation ratio.}
\label{fig:Cylinders2D_CDsD}
\end{figure}

The lift and drag coefficients for both the upstream and downstream cylinders are obtained for $s=4.5$ and are shown in Figure \ref{fig:Cylinders2Dvalidation}, where good agreement is observed comparing with reference data \cite{ding2007numerical}. The drag coefficients of the two cylinders are obtained by integrating the pressure and shear stress distributions on the surface and then are averaged for $500$ convective times. The time-averaged drag coefficient, $\overline{c_d}$, is plotted for different values of $s$ in Figure \ref{fig:Cylinders2D_CDsD}, and shows a similar trend to Igarashi \cite{igarashi1981characteristics}. The time-averaged drag coefficient of the upstream cylinder, $\overline{c_d}_1$, decreases gradually by increasing the cylinder spacing, $s$, and increases stepwise for $s/D>3$. On the other hand, the time-averaged drag coefficient of the downstream cylinder, $\overline{c_d}_2$, is negative for $s/D<3$ acting as a thrust force. $\overline{c_d}_2$ increases as the downstream cylinder is placed further away from the upstream cylinder, and a sudden increase occurs for $s/D>3$. 

\subsection{Optimization}

In this study, the distance between the two cylinders, $s$, and the ratio between the diameters of the cylinders, $r$, are the design variables, $\pmb{\mathcal{X}} = [s/D, r]$. The objective function is $\mathcal{F} = p^\prime_{rms}$ at $2D$ above the upstream cylinder, depicted in Figure \ref{fig:CylindersObserver}. Considering that the main objective of this study is to demonstrate the optimization capabilities of MADS, a loud initial design has been selected to evaluate the feasibility of optimizing towards a quieter configuration.

\subsubsection{Results and Discussion}

\begin{figure}
\centering
\begin{subfigure}{0.7\textwidth}
\includegraphics[width=\textwidth]{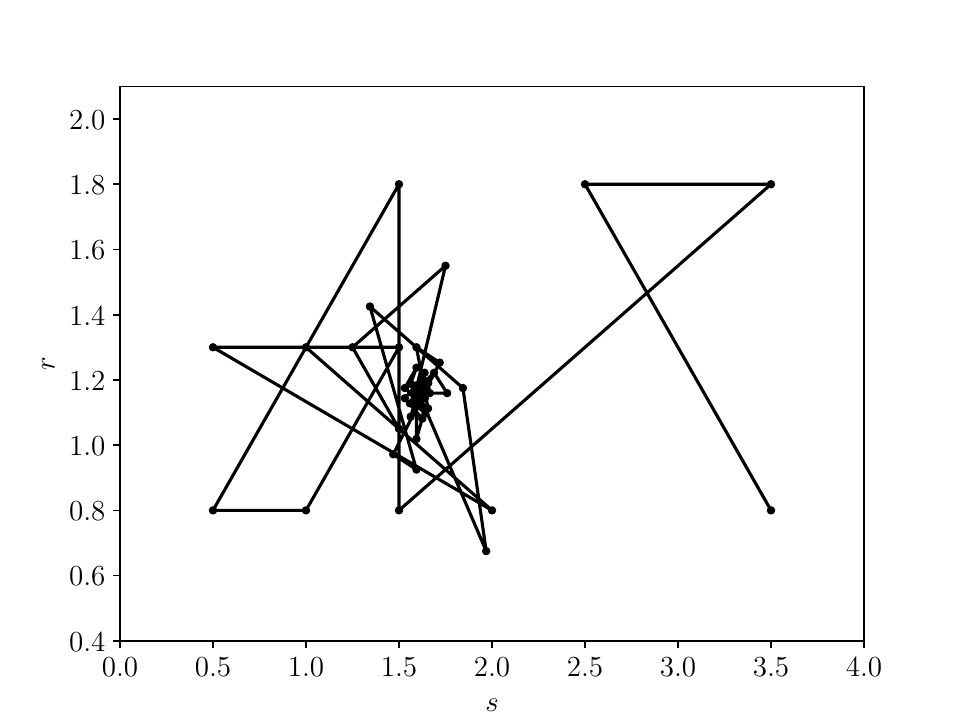}
\subcaption{The design space.}
\label{figure_cylinder_optimization_design_space}
\end{subfigure}
\begin{subfigure}{0.7\textwidth}
\includegraphics[width=\textwidth]{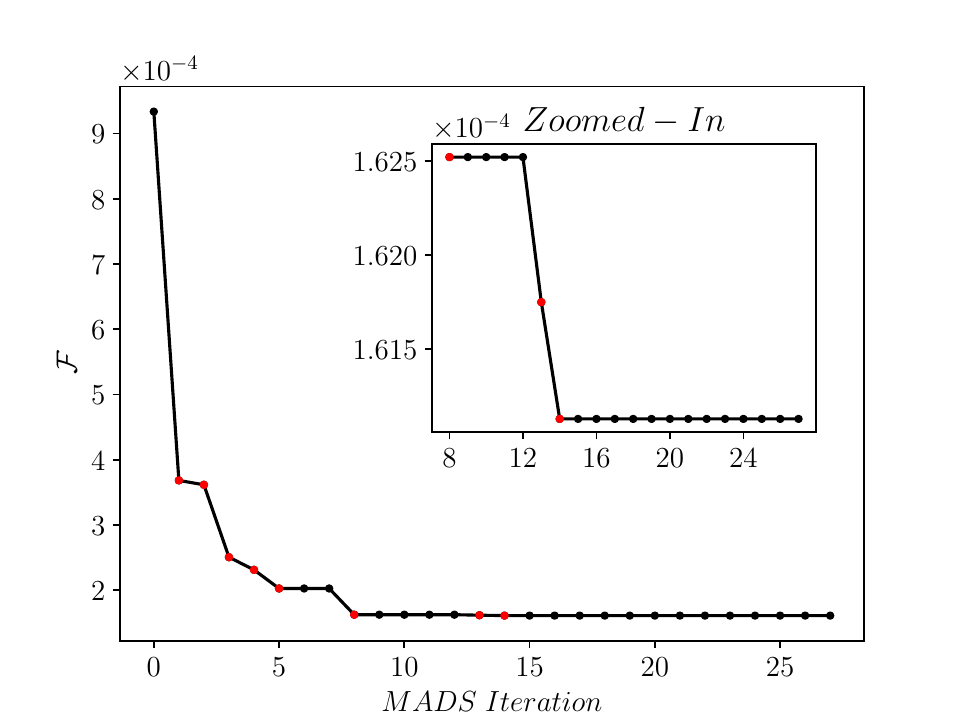}
\subcaption{The objective function convergence with the new incumbent designs highlighted in red.}
\label{figure_cylinder_optimization_obj}
\end{subfigure}
\caption{The design space and objective function convergence for the tandem cylinders.}
\label{figure_cylinder_optimization}
\end{figure}

\begin{figure}
\centering
\begin{subfigure}{\textwidth}
\centering
\begin{subfigure}{0.45\textwidth}
\includegraphics[width=\textwidth]{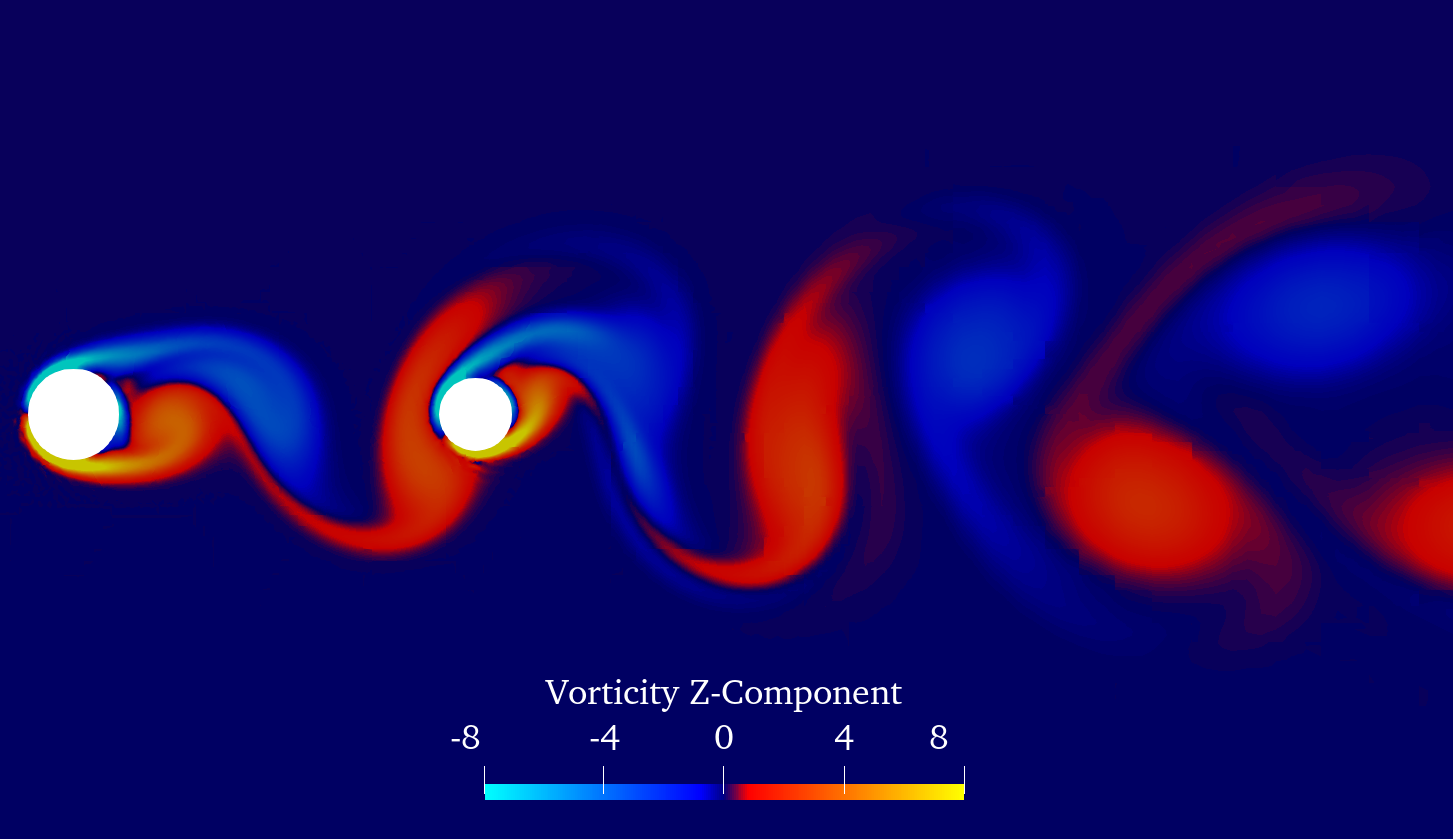}
\end{subfigure}
\begin{subfigure}{0.45\textwidth}
\includegraphics[width=\textwidth]{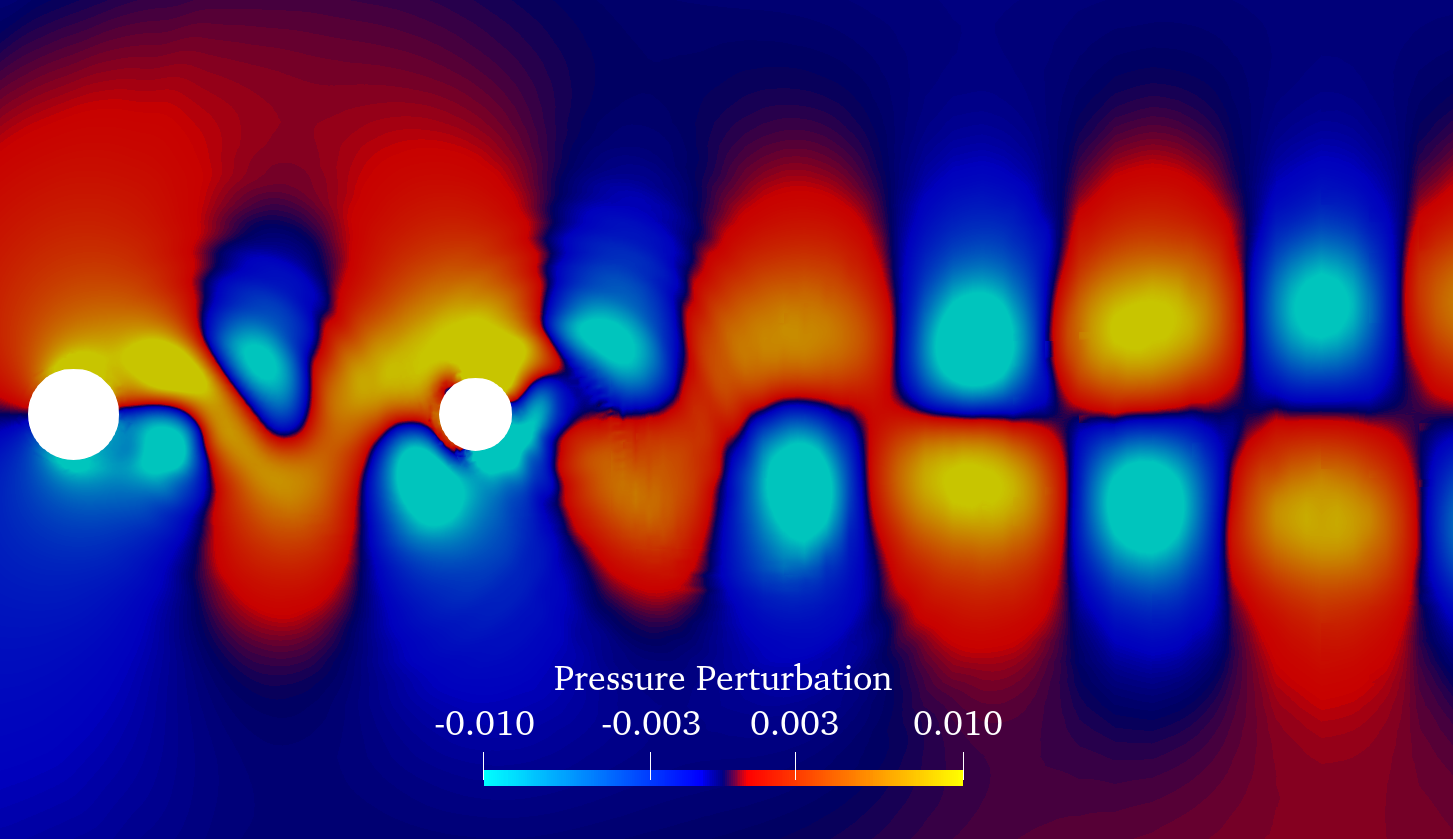}
\end{subfigure}
\subcaption{The initial tandem cylinder design.}
\label{fig:Cylinders2DinitialDesign}
\end{subfigure}
\begin{subfigure}{\textwidth}
\centering
\begin{subfigure}{0.45\textwidth}
\includegraphics[width=\textwidth]{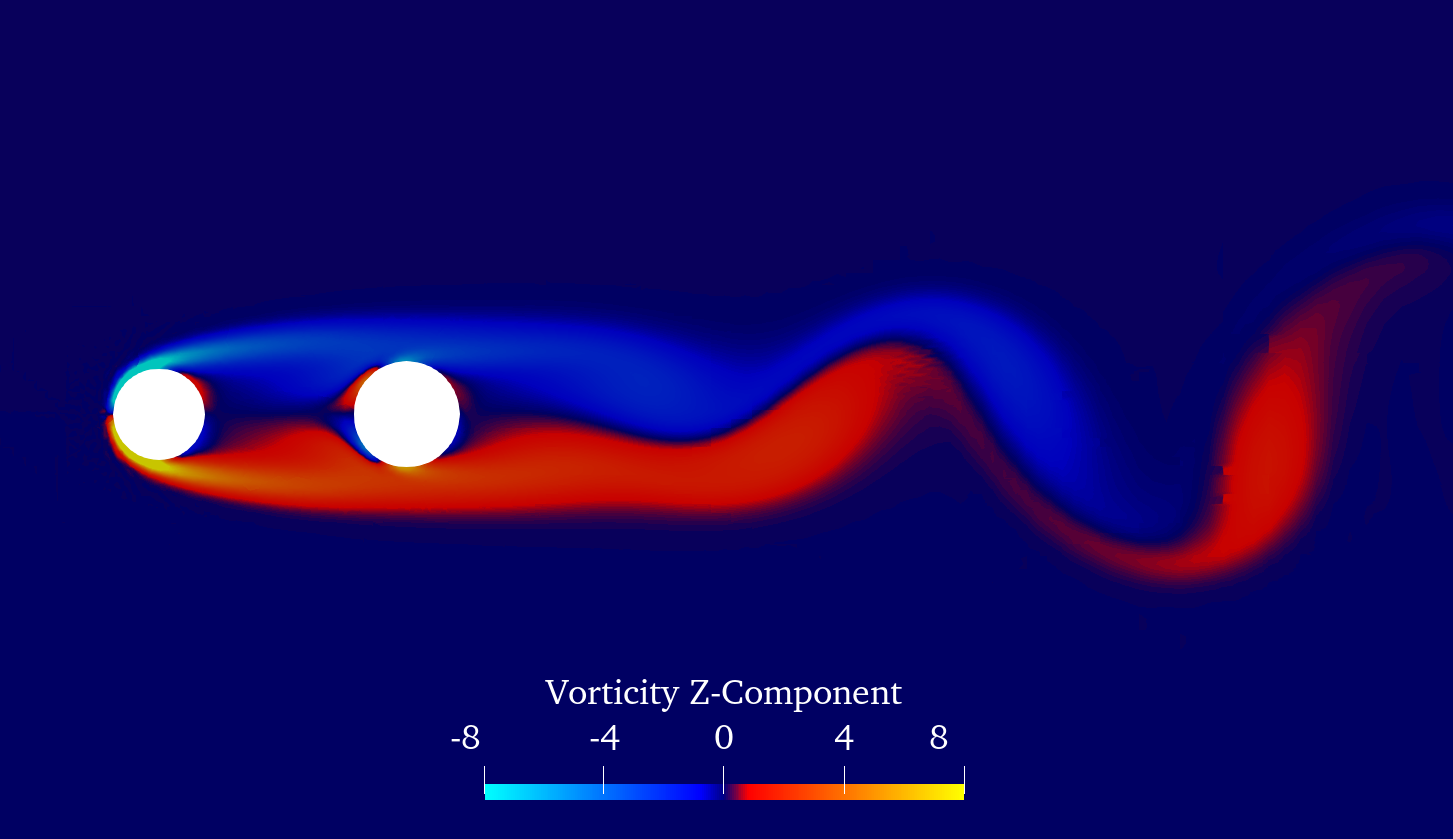}
\end{subfigure}
\begin{subfigure}{0.45\textwidth}
\includegraphics[width=\textwidth]{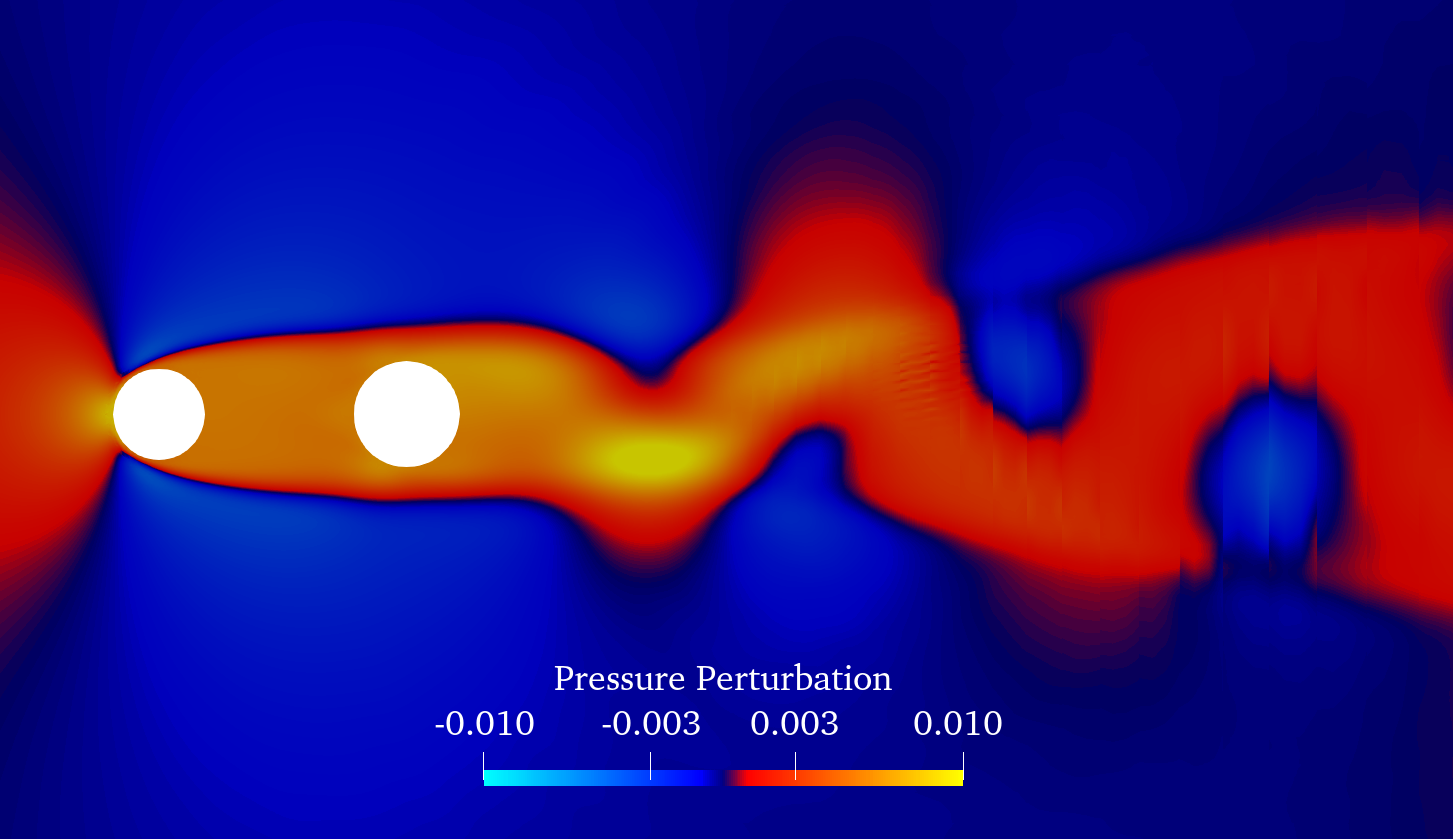}
\end{subfigure}
\subcaption{The optimum tandem cylinder design.}
\label{fig:Cylinders2DoptimumDesign}
\end{subfigure}
\caption{The z-component of the vorticity and pressure perturbation snapshots at $t_c=2000$.}
\label{fig:Cylinders2Doptimizations}
\end{figure}

The optimization problem converges after $27$ MADS iterations, including $70$ objective function evaluations. The design space and objective function convergence are shown in Figure \ref{figure_cylinder_optimization}, where the optimum design is found as $(s/D,r)=(1.6301,1.1594)$. The optimization procedure has covered a wide range of design variables, as shown in Figure \ref{figure_cylinder_optimization_design_space}. Instantaneous vorticity contours and acoustic fields are shown for the initial design and the optimum design, in Figure \ref{fig:Cylinders2Doptimizations}, at $t_c = 2000$. The overall sound pressure level of the initial design at the observer, $2D$ above the upstream cylinder, is $136.3~dB$, which reduces to $119.8~dB$ for the optimized configuration. Thus, a $16.5~dB$ decrease in overall SPL is achieved.

\section{NACA0012 Airfoil}
\label{sec:NACA0012}

The aerodynamic characteristics of the NACA0012 airfoil have been extensively studied through experiments \cite{paterson1973vortex, longhouse1977vortex, arbey1983noise, nash1999boundary} and CFD simulations \cite{atobe3flow, ikeda2012direct}. This airfoil has a relatively high maximum lift coefficient, which makes it suitable for use in low-speed applications such as general aviation, Unmanned Aerial Vehicles (UAVs), and Micro Aerial Vehicles (MAVs). At low Reynolds numbers, less than $Re=10^5$, the boundary layer is laminar. In general, two different types of acoustic spectra are observed in flow past a laminar airfoil, depending on the Reynolds number and angle of attack. First, a typical tone noise phenomenon, i.e., a broadband contribution with a dominant frequency along with equidistant frequency tones, and second, a simple broadband spectrum \cite{desquesnes2007numerical}. In the first type, the sequence of regularly spaced discrete frequency tones is due to the emergence of a separation bubble on the pressure surface close to the trailing edge \cite{desquesnes2007numerical}. On the pressure side, the hydrodynamics fluctuations are coherent in the spanwise direction \cite{paterson1973vortex}. Thus, it can be assumed that the governing mechanism of tonal noise is essentially two-dimensional \cite{desquesnes2007numerical}. 

The study of airfoil noise dates back to the 1970s when several experimental studies showed that discrete tones are emitted from isolated airfoils \cite{clark1971radiation, hersh1971aerodynamic}, and other studies focused on understanding this phenomenon \cite{paterson1973vortex, tam1974discrete, arbey1983noise}. The shape of the airfoil is optimized for noise reduction of the high-lift devices \cite{marsden2001shape}, laminar flow trailing edge \cite{marsden2002optimal, marsden2004optimal}, and turbulent flow trailing edge \cite{jouhaud2007surrogate, marsden2007trailing}. This study examines the laminar flow trailing edge and the aeroacoustic shape optimization of the NACA0012 airfoil at a low Reynolds number, $Re=10,000$, which is the operating regime for MAVs.

\subsection{Validation} 

In this section, flow over a two-dimensional NACA0012 airfoil is validated. A grid-resolution-independence study is performed for the time-averaged lift and drag coefficients, and the overall sound pressure level at an observer located a unit chord length below the trailing edge. The time-averaged lift coefficient is compared with reference DNS data \cite{ikeda2012numerical} to validate the simulation.

\subsubsection{Computational Details}

\begin{figure}
\centering
\begin{subfigure}{0.45\textwidth}
\centering
\includegraphics[width=\textwidth]{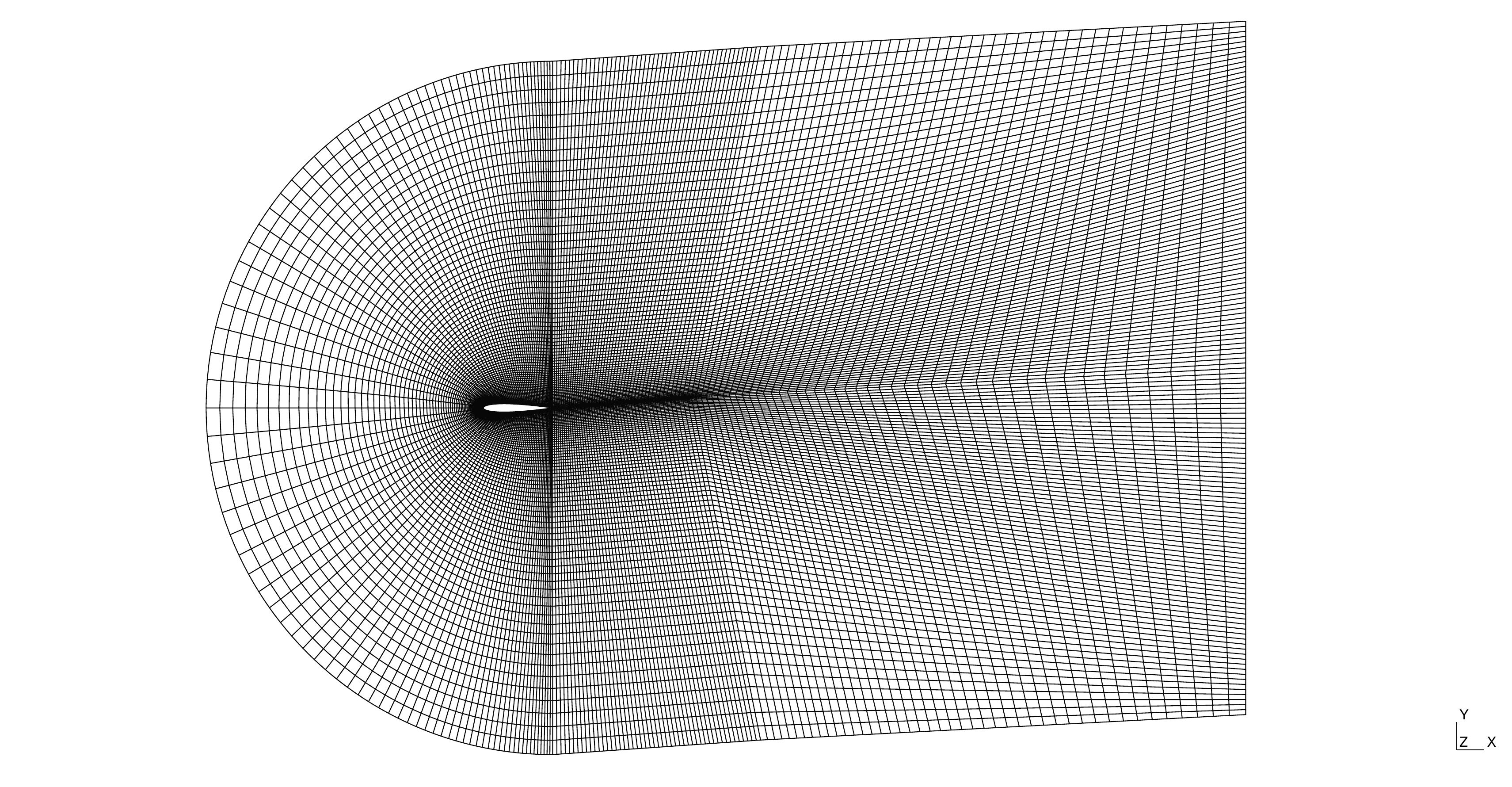}
\subcaption{The computational domain.}
\end{subfigure}
\begin{subfigure}{0.45\textwidth}
\centering
\includegraphics[width=\textwidth]{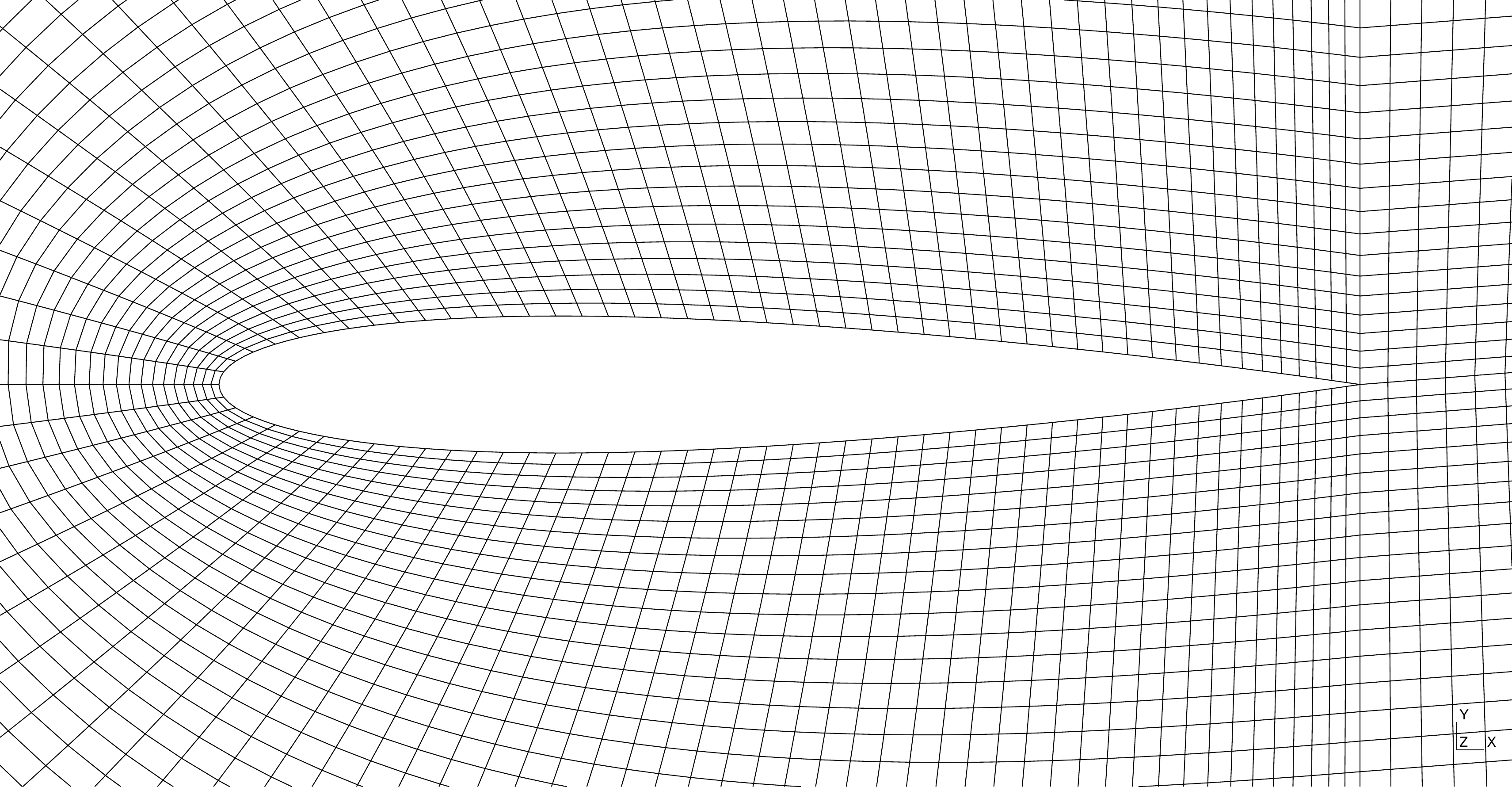}
\subcaption{The vicinity of the airfoil.}
\end{subfigure}
\caption{The computational grid for NACA0012 airfoil at $\alpha = 3^\circ$.}
\label{fig:Airfoil2Dmesh}
\end{figure}

The computational grid consists of $19,596$ quadrangular elements, depicted in Figure \ref{fig:Airfoil2Dmesh}. The domain is extended to $5c$ in the $y$-direction and to $10c$ in the $x$-direction, where $c=1$ is the chord length of the airfoil. The stretching ratio is kept below $5\%$ everywhere in the domain.
The elements in the wake region are inclined at the angle of attack to capture the vortices behind the trailing edge. The computational domain is shown in Figure \ref{fig:Airfoil2Dmesh}. The Reynolds number for this study is $Re=10,000$, the inflow Mach number is $M_\infty=0.2$, the angle of attack is $3$ degrees, and the Prandtl number is $Pr=0.71$. The simulation is run for $60$ convective times, and flow statistics are averaged for the last $20$ convective times. The second-order Paired Explicit Runge-Kutta (P-ERK) temporal scheme \cite{vermeire2019paired, vermeire2023embedded} is used to advance the solution in time. 

\begin{figure}
\centering
\begin{subfigure}{0.45\textwidth}
\centering
\includegraphics[width=\textwidth]{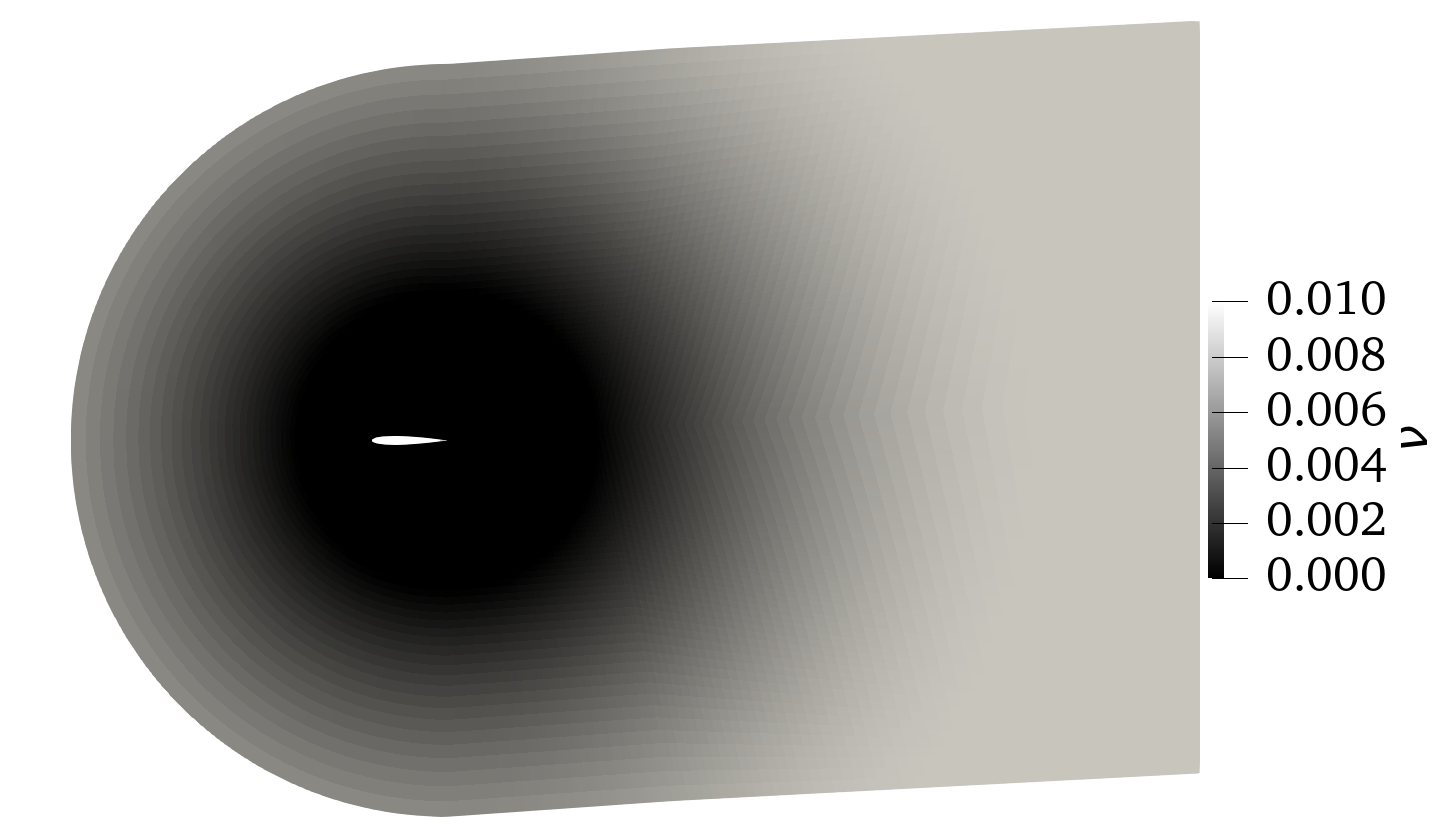}
\subcaption{The artificial diffusion.}
\label{fig:Airfoil2DBoundaryTreatmentsAV}
\end{subfigure}
\begin{subfigure}{0.45\textwidth}
\centering
\includegraphics[width=\textwidth]{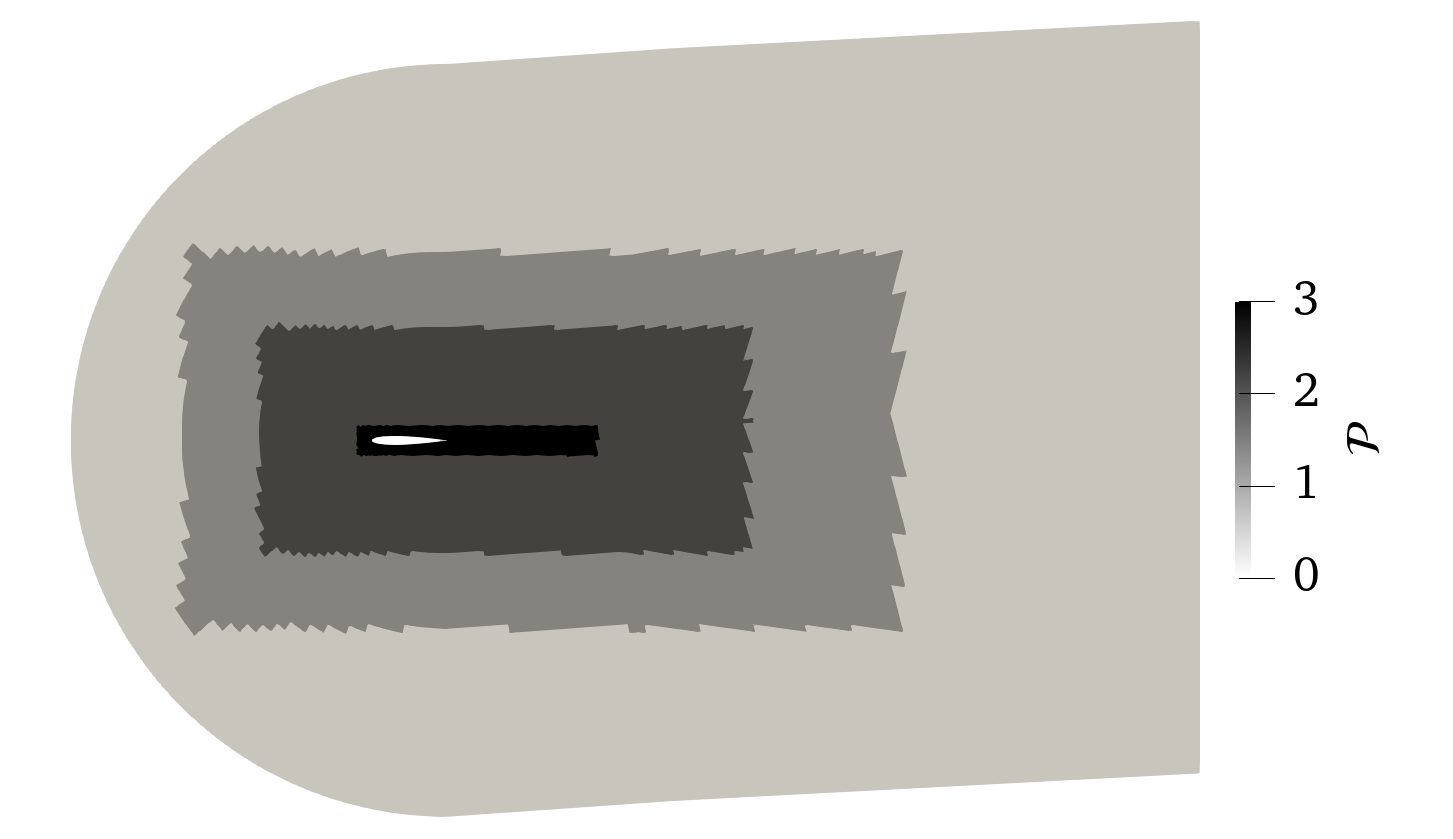}
\subcaption{The polynomial distribution.}
\label{fig:Airfoil2DBoundaryTreatmentsPadaptation}
\end{subfigure}
\caption{The boundary treatments.}
\label{fig:Airfoil2DBoundaryTreatments}
\end{figure}

Vortices leaving the computational domain can generate non-physical acoustic wave reflections off the boundaries, contaminating the solution. Thus, the strength of such vortices must be decreased to eliminate the acoustic wave reflections off the boundaries. The addition of artificial diffusion and variable solution polynomial degrees are used in this study, shown in Figure \ref{fig:Airfoil2DBoundaryTreatments}. Artificial diffusion is applied beyond a circle with a radius of $2c$ centered at the trailing edge. Its magnitude increases to a maximum of $0.01$ and a radius of $8c$ using a sinusoidal function. The solution polynomial distribution is shown in Figure \ref{fig:Airfoil2DBoundaryTreatmentsPadaptation}, where in the vicinity of the airfoil $\mathcal{P}=3$ and it decreases to zero close to the boundaries.

\subsubsection{Results and Discussion}

\begin{figure}
\centering
\begin{subfigure}{0.45\textwidth}
\centering
\includegraphics[width=\textwidth]{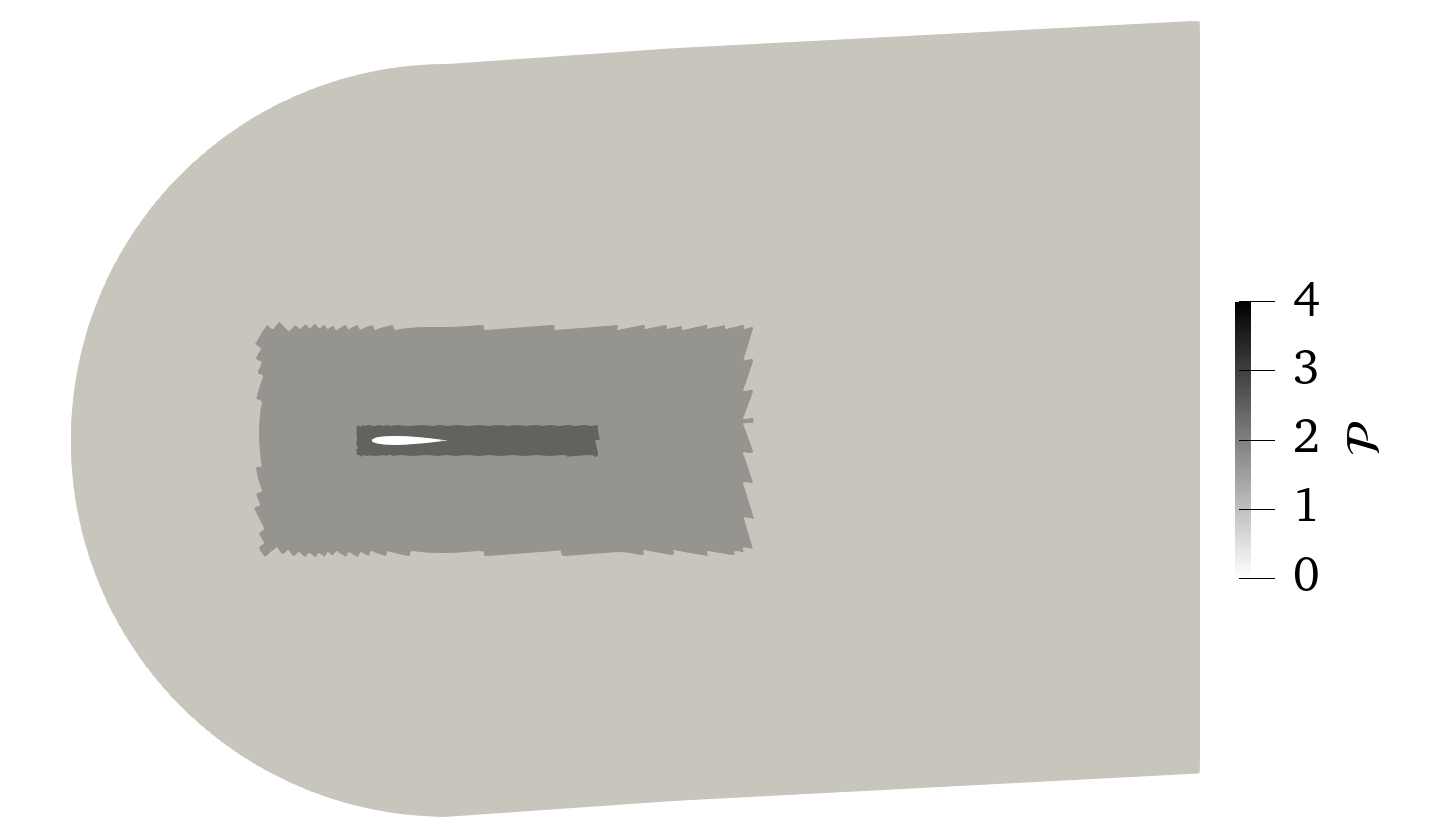}
\subcaption{Low resolution, $\mathcal{P}0-\mathcal{P}2$.}
\label{fig:Airfoil2DGridIndependenceP2}
\end{subfigure}
\begin{subfigure}{0.45\textwidth}
\centering
\includegraphics[width=\textwidth]{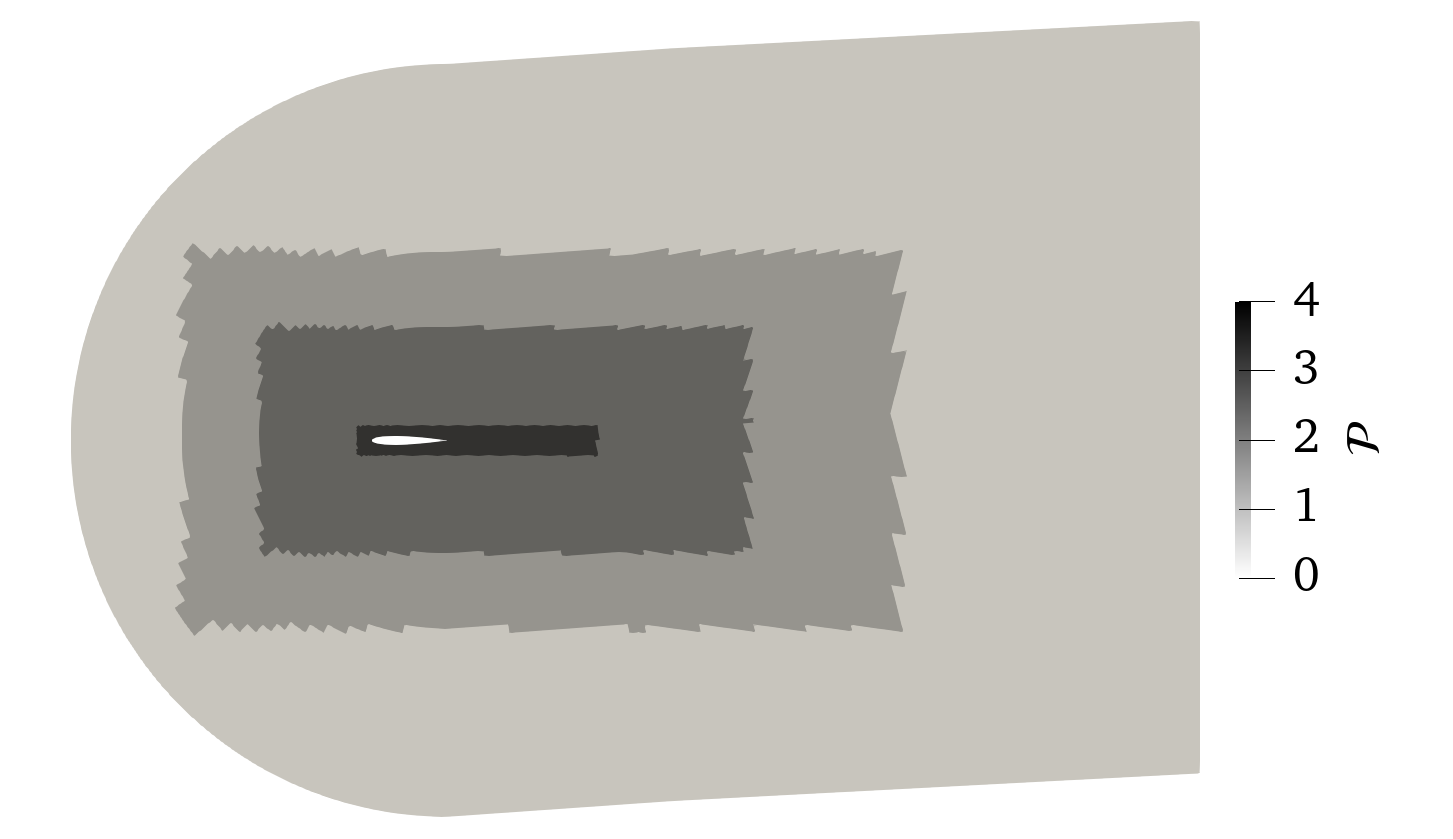}
\subcaption{Medium resolution, $\mathcal{P}0-\mathcal{P}3$.}
\label{fig:Airfoil2DGridIndependenceP3}
\end{subfigure}
\begin{subfigure}{0.45\textwidth}
\centering
\includegraphics[width=\textwidth]{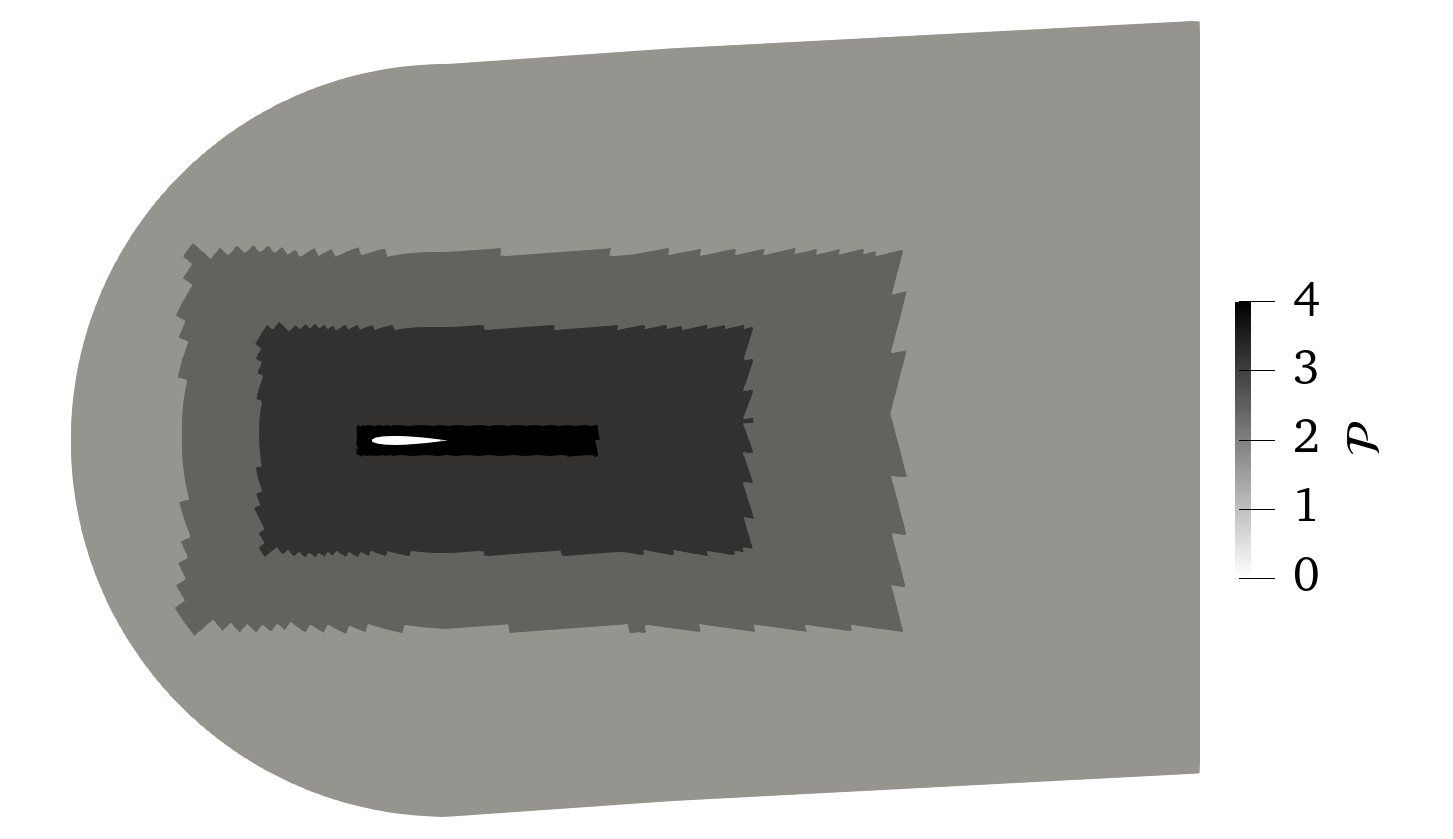}
\subcaption{High resolution, $\mathcal{P}1-\mathcal{P}4$.}
\label{fig:Airfoil2DGridIndependenceP4}
\end{subfigure}
\caption{The three different resolutions to study the grid resolution independency of the results.}
\label{fig:Airfoil2DGridIndependence}
\end{figure}

A different set of variable polynomial degrees is used to study the independence of the results to the grid resolution. Three different mesh resolutions are used with a maximum polynomial degree of $\mathcal{P}2$, $\mathcal{P}3$, and $\mathcal{P}4$, shown in Figure \ref{fig:Airfoil2DGridIndependence}. 

The time-averaged lift and drag coefficients are computed along with the overall sound pressure level at the observer located a unit chord length below the trailing edge and compared using three different grid resolutions, shown in Table \ref{table:Airfoil2DClCdSPL}. The time-averaged lift coefficient differs by less than $0.4\%$ when the highest polynomial degree in the domain is $\mathcal{P}4$ compared to that of $\mathcal{P}3$, while the time-averaged drag coefficient remains the same by three significant digits. The time-averaged lift coefficient obtained via the $\mathcal{P}0 - \mathcal{P}3$ simulation agrees well with the DNS data \cite{ikeda2012numerical}. Furthermore, the overall sound pressure level difference between $\mathcal{P}0 - \mathcal{P}3$ and $\mathcal{P}1 - \mathcal{P}4$ simulations is $0.53~dB$ or $0.48\%$. Thus, it is concluded that the grid resolution for $\mathcal{P}0 - \mathcal{P}3$ simulation is sufficient for this problem.  

\begin{table}
\centering
\caption{Averaged lift and drag coefficients and the overall sound pressure level measured at a unit chord distance below the trailing edge of the baseline NACA0012.}
\begin{tabular}{cccc}
\hline
 & $\mathcal{P}0-\mathcal{P}2$ & $\mathcal{P}0-\mathcal{P}3$ & $\mathcal{P}1-\mathcal{P}4$ \\
\hline
$\overline{c_l}$ & $0.0877$ & $0.0886$ & $0.0889$ \\
$\overline{c_d}$ & $0.0448$ & $0.0447$ & $0.0447$ \\
$OASPL$ & $109.59~dB$ & $110.00~dB$ & $110.53~dB$ \\
\hline
\end{tabular}
\label{table:Airfoil2DClCdSPL}
\end{table}

The sensitivity of the time-averaged quantities to the averaging window is investigated by choosing two different averaging windows. The lift and drag coefficients are averaged over $20$ and $40$ convective time windows, shown in Table \ref{table:Airfoil2DClCdAvgWindow}. The difference between the time-averaged lift and drag coefficients for both averaging window lengths is negligible. Thus, the quantities are averaged over a $20$ convective time window.

\begin{table}
\centering
\caption{Averaging window sensitivity of the time-averaged quantities.}
\begin{tabular}{ccc}
\hline
 & $20~t_c$ & $40~t_c$ \\
\hline
$\overline{c_l}$ & $0.08857$ & $0.08863$ \\
$\overline{c_d}$ & $0.04472$ & $0.04473$ \\
\hline
\end{tabular}
\label{table:Airfoil2DClCdAvgWindow}
\end{table}

\begin{figure}
\centering
\includegraphics[width=0.7\textwidth]{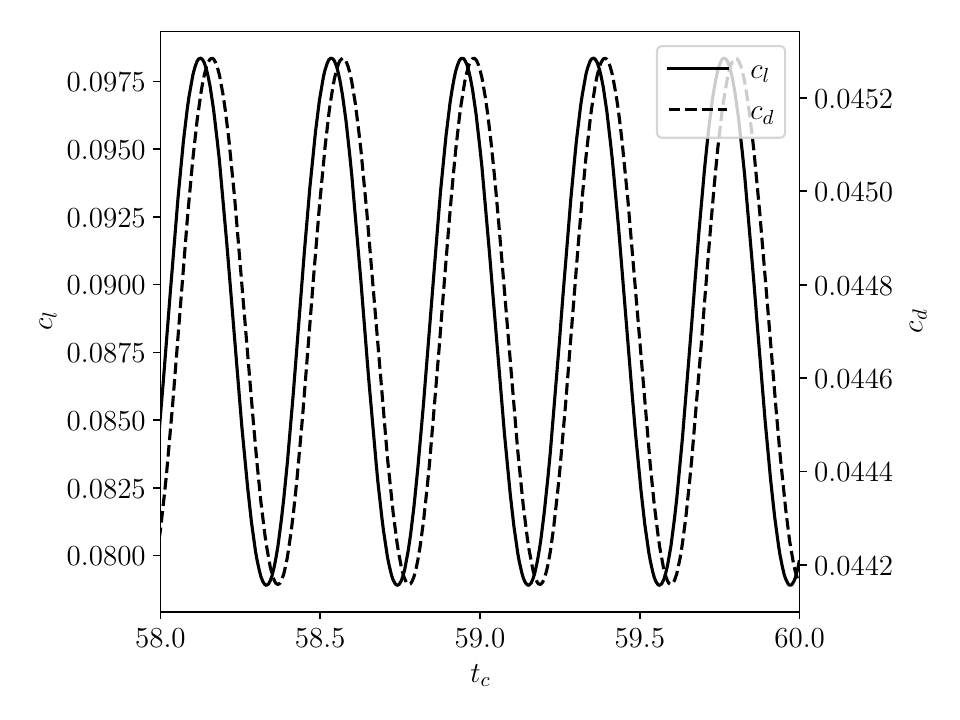}
\caption{Time-histories of lift and drag coefficients.}
\label{fig:Airfoil2DClCdTimeHistories}
\end{figure}

The time history of lift and drag coefficients are shown in Figure \ref{fig:Airfoil2DClCdTimeHistories} for the last two convective times. The periodic behavior of $c_l$ and $c_d$ is associated with the periodic vortex shedding at the trailing edge.

\noindent The pressure perturbation at $t_c=60$ is shown in Figure \ref{fig:Airfoil2DP2Optimization}. Acoustic waves are generated close to the trailing edge and propagate everywhere in the domain. There are no visible acoustic wave reflections off the boundaries, showing the effectiveness of boundary treatments used in this study. The amplitude of the pressure perturbations is higher in the wake region and behind the trailing edge where the vortices are shed and travel downstream. The addition of artificial viscosity, as shown in Figure \ref{fig:Airfoil2DBoundaryTreatmentsAV},  dampens these vortices and consequently reduces the amplitude of acoustic waves far from the trailing edge. 

\subsection{Optimization}

\begin{figure}
\centering
\includegraphics[width=0.8\textwidth]{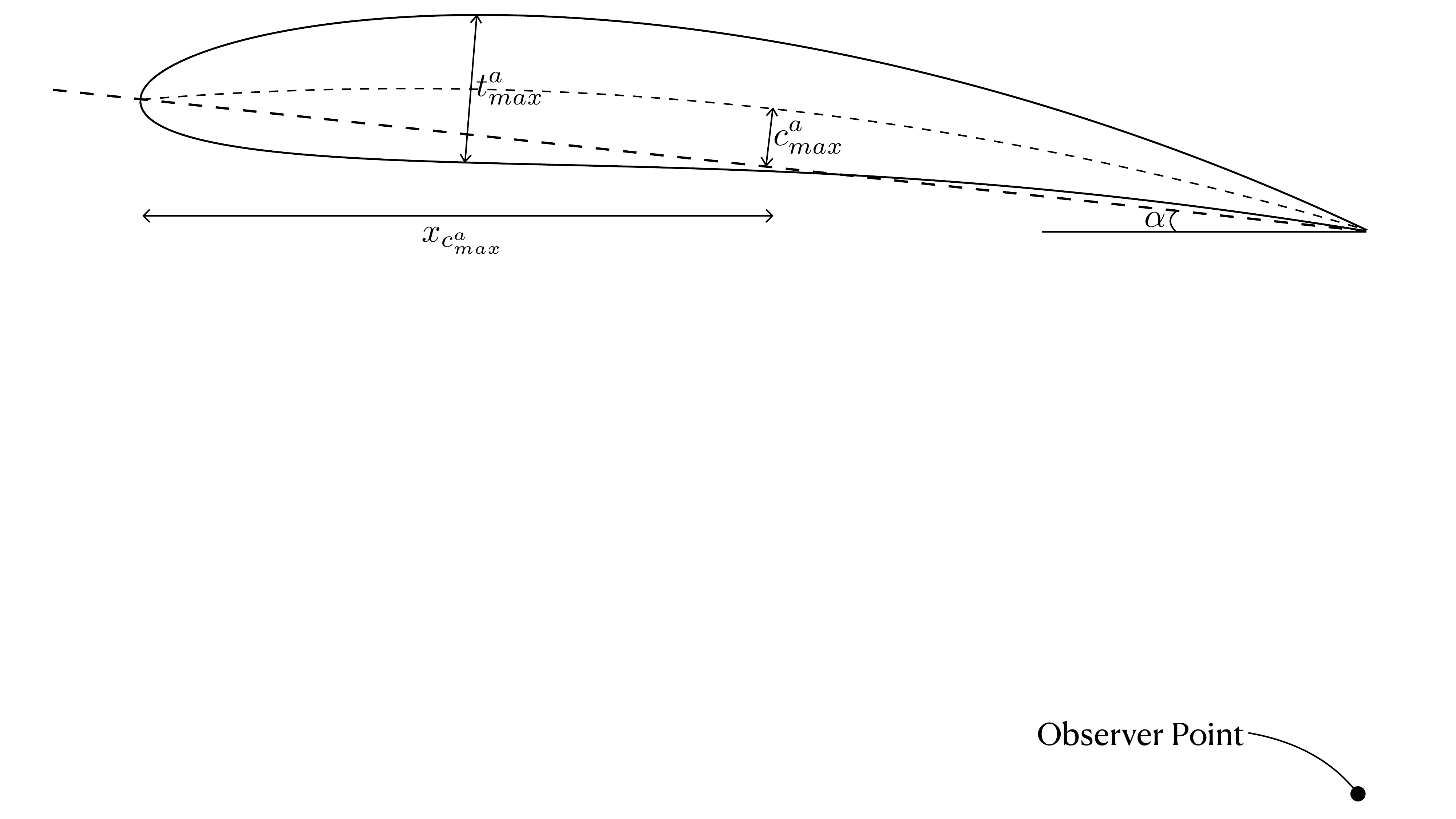}
\caption{The design variables and the observer point located at a unit chord length below the trailing edge for the two-dimensional NACA 4-digit airfoil.}
\label{fig:Airfoil2Dgeometry}
\end{figure}

In this section, the noise at an observer located at a unit chord length below the trailing edge is reduced. A total of four design parameters are chosen based on the NACA 4-digit airfoil series. The maximum camber, $c_{max}^{a}$, the distance of maximum camber from the airfoil leading edge, $x_{c_{max}^{a}}$, maximum thickness of the airfoil $t_{max}^{a}$, and the angle of attack, $\alpha$, are the four design parameters, i.e. $\pmb{\mathcal{X}} = [c_{max}^{a}, x_{c_{max}^{a}}, t_{max}^{a}, \alpha]$, depicted in Figure \ref{fig:Airfoil2Dgeometry}. The simulation is first run for $60$ convective times for each objective function evaluation. Then the time-averaged pressure is computed from $20t_c$ to $40t_c$, $\overline{p}_{20-40}$, and then from $40t_c$ to $60t_c$, $\overline{p}_{40-60}$. If the difference between $\overline{p}_{20-40}$ and $\overline{p}_{40-60}$ is above one percent, the simulation is run for $20$ more convective times. The simulation is run long enough so that the difference between two consecutive time-averaged pressure signals, over $20t_c$, is below one percent. 

\subsubsection{Results and Discussion}

The optimization procedure is run using a maximum polynomial degree of $\mathcal{P}3$, shown in Figure \ref{fig:Airfoil2DGridIndependenceP3}. The maximum camber range is set to $c_{max}^{a} \in [-10, 10]$ as a percentage of the chord, with the distance from the airfoil leading edge in the range of $x_{c_{max}^{a}} \in [2, 9]$ as a tenth of the chord. The maximum thickness of the airfoil is within the range of $t_{max}^{a} \in [8,16]$ as a percentage of the chord. Finally, the angle of attack varies from $-5^\circ$ to $5^\circ$. The objective function is defined as the overall sound pressure level at the observer with constraints on both the mean lift and mean drag coefficients. A quadratic penalty term is added to the objective function when the lift coefficient deviates from the baseline design, and an additional quadratic penalty term is added when the mean drag coefficient is above the baseline design. The objective function is defined as 

\begin{align}
&
\mathcal{F} = 
\begin{cases}
OASPL + \epsilon \left( \overline{c_l} - \overline{c_{l,baseline}} \right)^2 + \epsilon \left( \overline{c_d} - \overline{c_{d,baseline}} \right)^2 & \overline{c_d} > \overline{c_{d,baseline}} \\
OASPL + \epsilon \left( \overline{c_l} - \overline{c_{l,baseline}} \right)^2 & \overline{c_d} \leq \overline{c_{d,baseline}}  \\
\end{cases}
&
,
\end{align}
where the constant $\epsilon$ is set to $400,000$ to compensate for the order of magnitude difference in $OASPL$ and $\overline{c_l}$ and $\overline{c_d}$. The defined objective function minimizes the overall sound pressure level while maintaining the mean lift coefficient, and ensures the optimized airfoil has a similar or lower mean drag coefficient. 

\begin{figure}
\centering
\begin{subfigure}{0.75\textwidth}
\centering
\includegraphics[width=\textwidth]{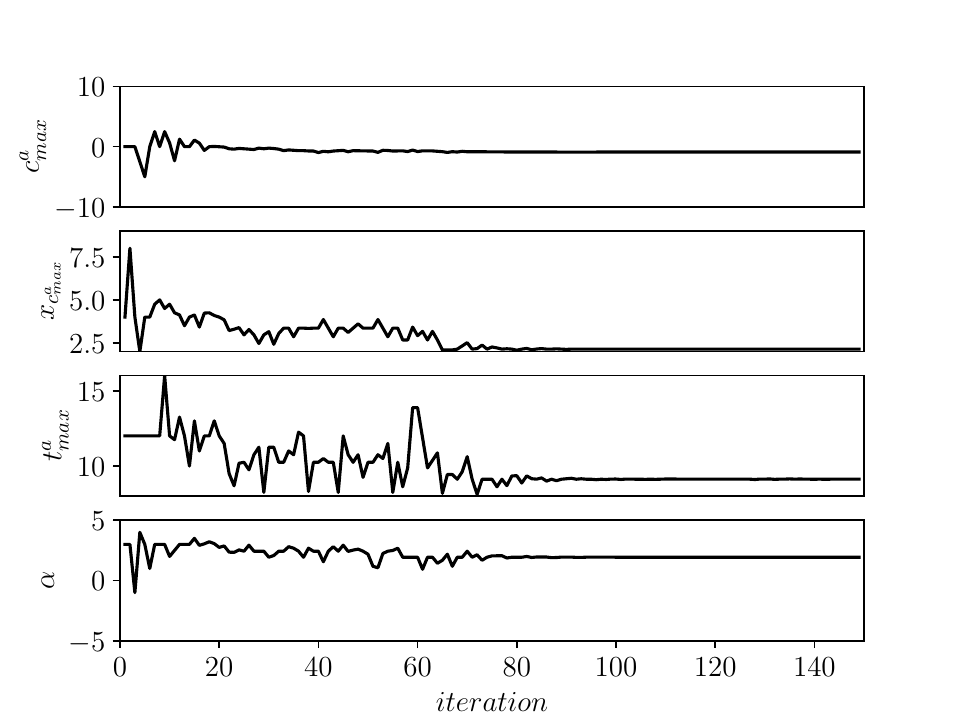}
\subcaption{The design space.}
\label{fig:Airfoil2DdesignSpaceP2Simulation}
\end{subfigure}
\begin{subfigure}{0.75\textwidth}
\centering
\includegraphics[width=\textwidth]{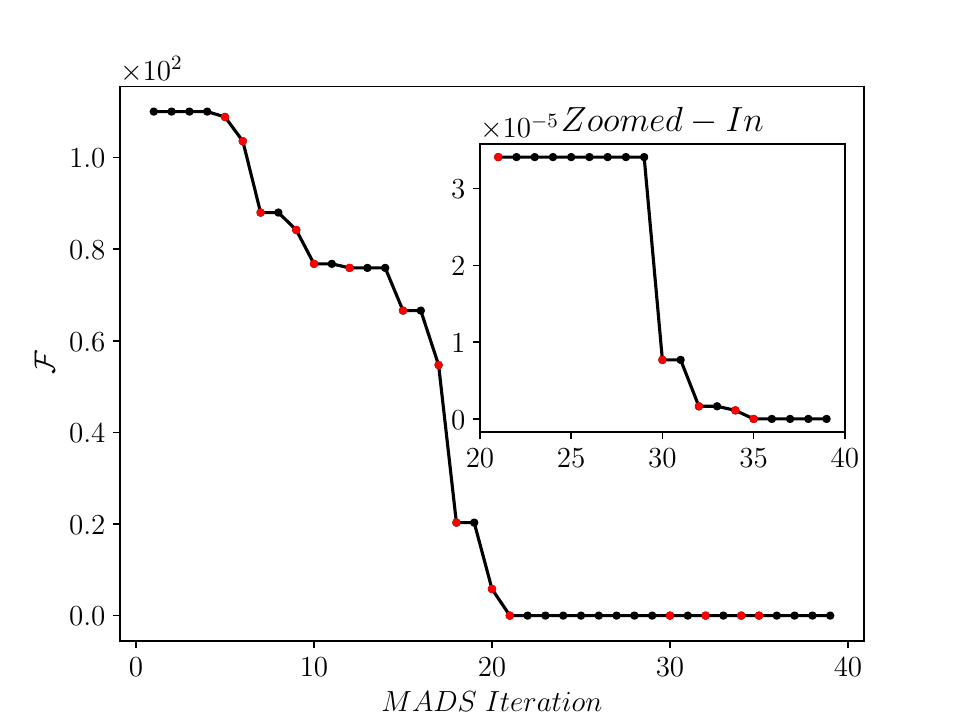}
\subcaption{The objective function convergence with the new incumbent design highlighted in red.}
\label{fig:Airfoil2DobjectiveFunctionP2Simulation}
\end{subfigure}
\caption{The design space and objective function convergence for $\mathcal{P}3$ optimization of the NACA 4-digit airfoil.}
\label{fig:Airfoil2DdesignSpaceObjFunctionP2Simulation}
\end{figure}

\begin{figure}
\centering

\begin{subfigure}{\textwidth}
\centering
\begin{subfigure}{0.45\textwidth}
\centering
\includegraphics[width=\textwidth]{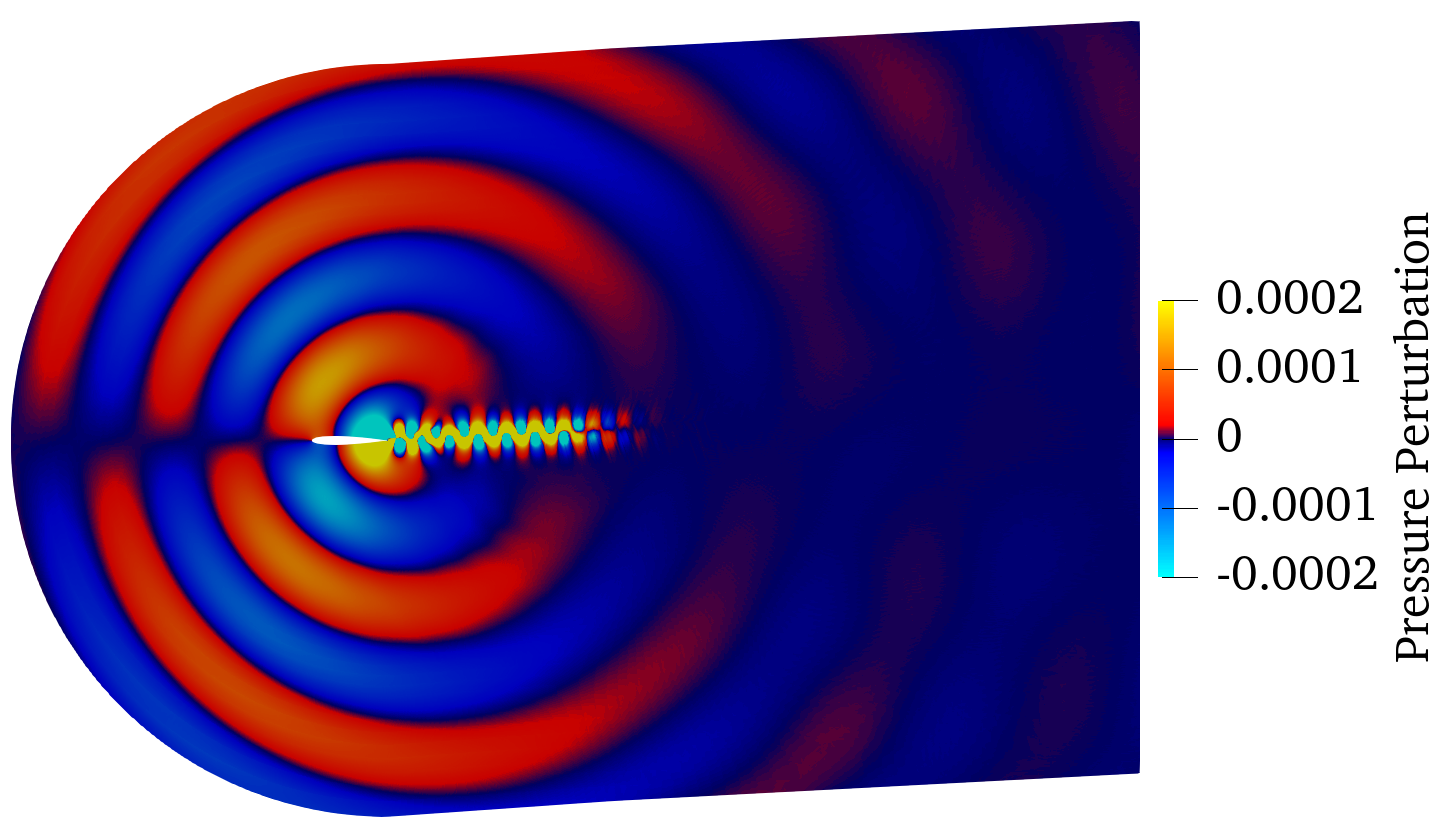}
\label{fig:Airfoil2DacousticPressureBaseline}
\end{subfigure}
\begin{subfigure}{0.45\textwidth}
\centering
\includegraphics[width=\textwidth]{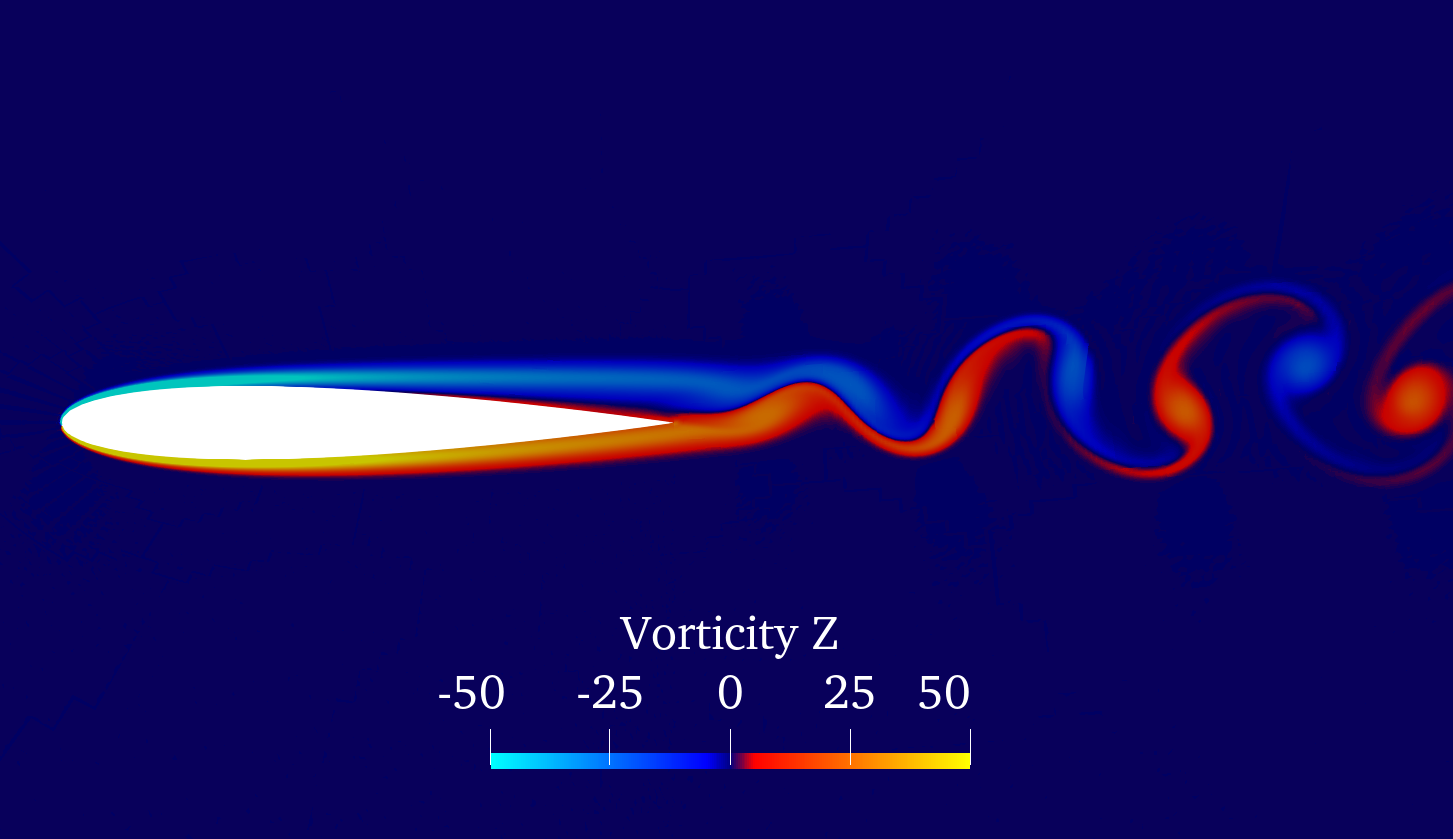}
\label{fig:Airfoil2DvorticityBaseline}
\end{subfigure}
\subcaption{The baseline design.}
\label{fig:Airfoil2DP2Optimization}
\end{subfigure}

\begin{subfigure}{\textwidth}
\centering
\begin{subfigure}{0.45\textwidth}
\centering
\includegraphics[width=\textwidth]{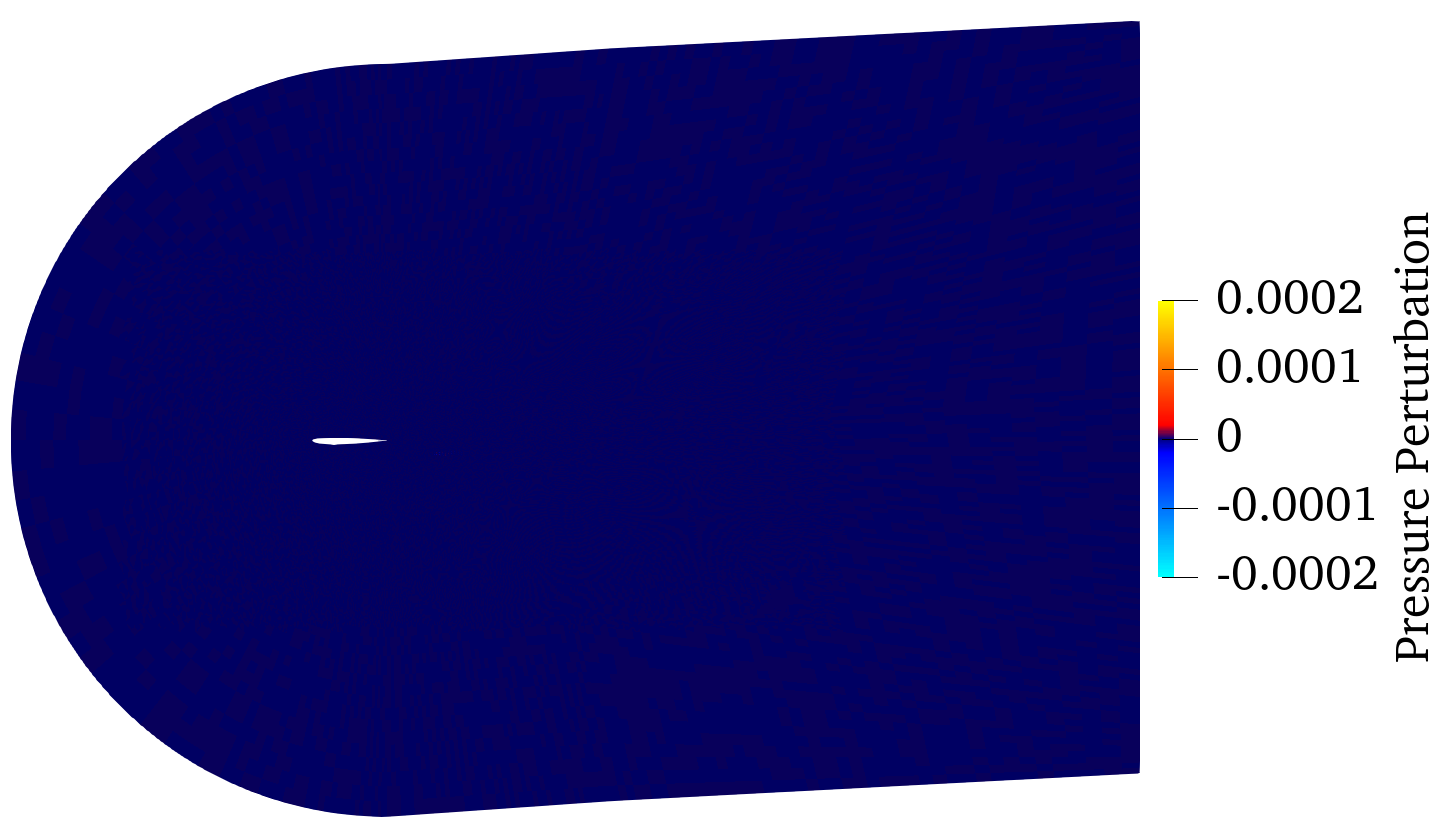}
\label{fig:Airfoil2DacousticPressureP3Optimization}
\end{subfigure}
\begin{subfigure}{0.45\textwidth}
\centering
\includegraphics[width=\textwidth]{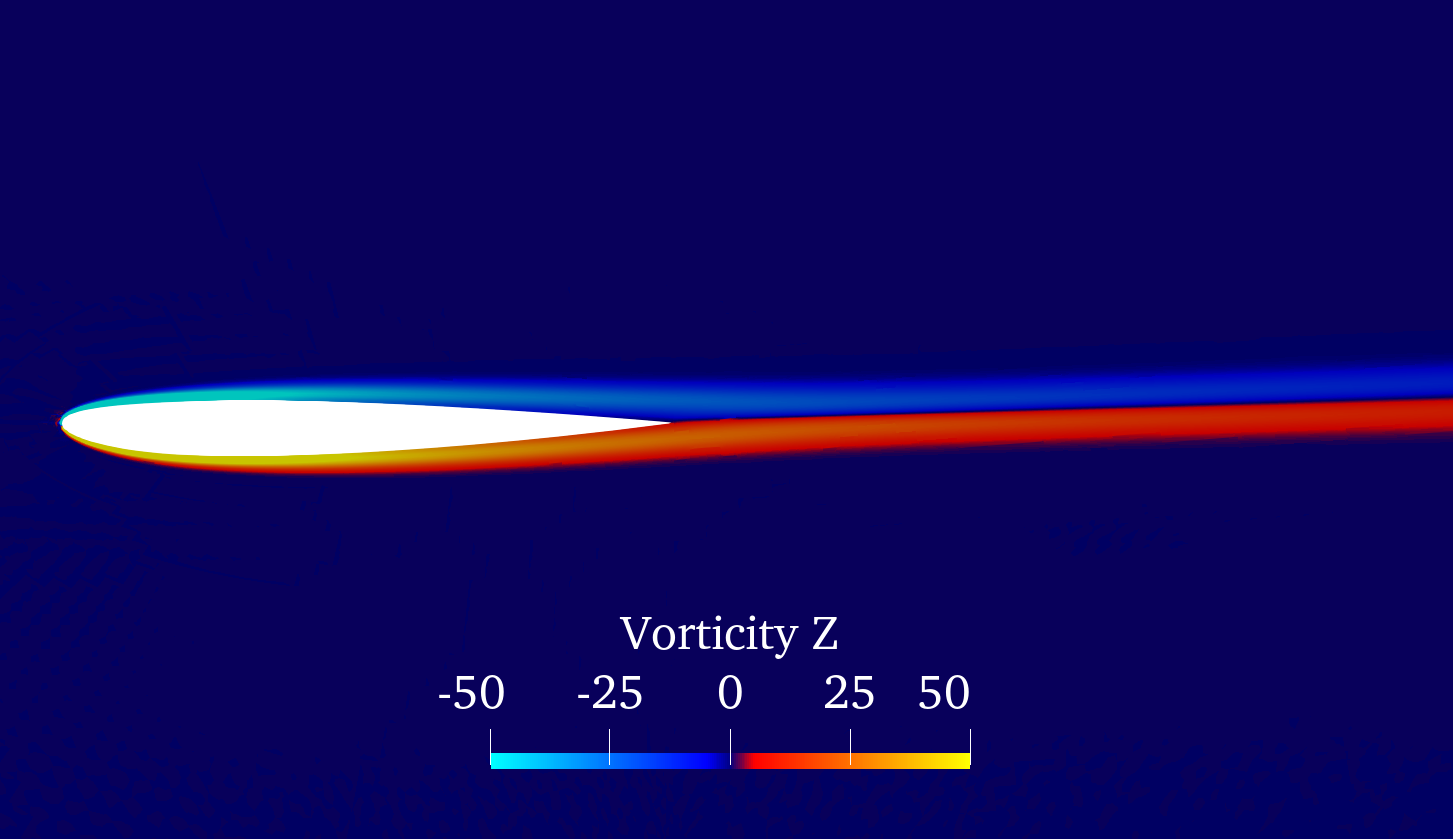}
\label{fig:Airfoil2DvorticityP3Optimization}
\end{subfigure}
\subcaption{The optimum design.}
\end{subfigure}

\caption{The pressure perturbation and vorticity in the z-direction for the baseline and $\mathcal{P}3$ optimization designs of NACA 4-digit airfoil at $t_c=60$.}
\label{fig:Airfoil2Doptimization}
\end{figure}

This optimization procedure converges after $39$ MADS iterations, consisting of $149$ objective function evaluations. The design space and the convergence of the objective function are shown in Figure \ref{fig:Airfoil2DdesignSpaceObjFunctionP2Simulation}. The optimal airfoil design has a maximum camber of $c_{max}^{a} = -0.8944$ percent of the chord, at a $2.1428$ tenth of the chord distance from the leading edge, with a thickness of $t_{max}^{a} = 9.1309$ percent of the chord, at an angle of attack of $\alpha = 1.9350$ degrees. The optimized airfoil is silent with $OASPL=0~dB$, maintains an unchanged mean lift coefficient of $\overline{c_l} = 0.0886$, and achieves a reduced mean drag coefficient by $24.95\%$ to $\overline{c_d} = 0.0348$. And, finally, the pressure perturbation and z-component of vorticity are shown in Figure \ref{fig:Airfoil2Doptimization} for the baseline and optimum designs. In the baseline design, the flow is attached to the airfoil on the pressure side, and flow instability occurs on the suction side. A periodic vortex shedding takes place as the flow passes over the trailing edge, resulting in acoustic wave generation. However, in the optimum design, the flow instability is eliminated resulting in a silent airfoil.

\section{Conclusions}
\label{sec:Conclusions}

This paper demonstrates the efficacy of using the high-order flux reconstruction method, direct acoustic computation, and mesh adaptive direct search optimization technique for aeroacoustic shape optimization. Three case studies were presented, demonstrating that significant noise reduction could be achieved by optimizing the geometry of two-dimensional objects at low Reynolds numbers. These findings have practical applications in various industries, including aerospace, automotive, and wind turbine design, where noise reduction is crucial. We note that the simulations undertaken here are relatively inexpensive two-dimensional problems. The computational cost of performing three-dimensional scale-resolving simulations, such as LES or DNS, are significantly higher than those in previous aerodynamic optimization studies due to domain size and resolution requirements \cite{karbasian2022gradient, caros2023optimization}.  Additionally, we propose the adaptivity and novel time stepping algorithms are two possible methods for reducing the computational cost of future LES or DNS simulations \cite{tugnoli2017locally, hedayati2021optimal, vermeire2023embedded}.  Despite this, the current work demonstrates the feasibility of the approach, and lays a foundation for future LES or DNS aeroacoustic optimization studies, provided appropriate computational resources.  Future research could extend this approach to more complex geometries, higher dimensions, and higher Reynolds numbers to explore the potential limits of the optimization technique and broaden its applications. While the baseline designs were validated against reference numerical simulations, there are currently no reference datasets for the newly optimized designs. In future studies, it would be prudent to conduct experimental studies using anechoic wind tunnels on the optimized configurations to validate the numerical results presented herein.

\section*{Data Statement}

Data relating to the results in this manuscript can be downloaded from the publication’s website under a CC-BY-NC-ND 4.0 license.

\section*{CRediT authorship contribution statement}

\textbf{Mohsen Hamedi:} Conceptualization; Data curation; Formal analysis; Investigation; Methodology; Software; Validation; Visualization; Writing - original draft. 
\textbf{Brian Vermeire:} Conceptualization; Funding acquisition; Investigation; Methodology; Project administration; Resources; Software; Supervision; Writing - review \& editing.

\section*{Declaration of competing interest}

The authors declare that they have no known competing financial interests or personal relationships that could have appeared to influence the work reported in this paper.

\section*{Acknowledgements}

The authors acknowledge support from the Natural Sciences and Engineering Research Council of Canada (NSERC) [RGPIN-2017-06773] and the Fonds de recherche du Québec (FRQNT) via the nouveaux chercheurs program.

\bibliography{manuscript}

\end{document}